\documentclass[12pt]{article}
\usepackage{amsthm}
\usepackage{eurosym}
\usepackage{amsfonts}
\usepackage{amssymb,amsmath}
\usepackage[dvips]{color}
\usepackage{amscd}
\usepackage{tikz}
\usepackage{mathtools}
\usepackage{graphicx}
\usepackage{caption}
\usepackage{subcaption}

\newcommand{\xdownarrow}[1]{ {\left\downarrow\vbox to #1{}\right.\kern-\nulldelimiterspace} }
\usetikzlibrary{decorations.pathreplacing}

\usetikzlibrary{patterns}

\def\and{\mathrm{and}}

\newcommand{\ee}{\end{equation}}
\newcommand{\bea}{\begin{eqnarray}}
\newcommand{\eea}{\end{eqnarray}}
\newcommand{\beas}{\begin{eqnarray*}}
\newcommand{\eeas}{\end{eqnarray*}}
\newcommand{\ba}{\begin{array}}
\newcommand{\ea}{\end{array}}

\newcommand{\nbox}{{\,\lower0.9pt\vbox{\hrule \hbox{\vrule height 0.2 cm \hskip 0.19 cm \vrule height 0.2 cm}\hrule}\,}}

\def\href#1#2{#2}
\theoremstyle{plain}

\begin{document}


\begin{titlepage}
\hfill
\vbox{
    \halign{#\hfil         \cr
           } 
      }  

\hbox to \hsize{{}\hss \vtop{ \hbox{}

}}

%

\vspace*{20mm}

\begin{center}

{\Large \textbf{Coherent state operators, giant gravitons, and gauge-gravity correspondence \vspace{0.4cm}}}

{\normalsize \vspace{10mm} }

{\normalsize {Hai Lin}}

{\normalsize \vspace{10mm} }

{\small \emph{\textit{Shing-Tung Yau Center of Southeast University, Southeast University,\\
Nanjing 210096, China
}} }

{\normalsize \vspace{0.2cm} }

{\small \emph{\textit{School of Mathematics, Southeast University, Nanjing 211189, China
}} }

{\normalsize \vspace{0.2cm} }

{\small \emph{\textit{Yau Mathematical Sciences Center, Tsinghua University,
Beijing 100084, China
\\
}} }
{\normalsize \vspace{0.4cm} }
\end{center}


\begin{abstract}

{\normalsize \vspace{0.3cm} }

We generalize a construction of coherent state operators describing various giant graviton branes. We  enlarge the coherent state parameters, by including complementary coherent state parameters, to describe a system of dual giants and giants. One of the advantages of using complementary coherent state parameters is that they have rich sub-block structures that record different type of giant gravitons or wrapped branes. The coherent state parameters are further packaged into supermatrix, to construct special coherent states, which encode information of both giants and dual giants. We add strings onto the coherent state operators. The string-added states capture near-BPS states. Hence the constructions of BPS coherent state operators are also useful for analyzing near-BPS states. The coherent state representation, auxiliary integrals, as well as auxiliary susy integrals, facilitate the computations efficiently. We describe multi-matrix operators and BPS states, and unify some classes of operators. Finally, reduced coherent states as well as their fermionic counterparts are discussed.

\end{abstract}

\end{titlepage}

\vskip 1cm

\section{Introduction}

\label{sec_introduction}\vspace{0.02cm}

The gauge-gravity correspondence \cite%
{Maldacena:1997re,Gubser:1998bc,Witten:1998qj} is a nontrivial
correspondence between a quantum theory with gravity in the bulk and a
different quantum system on the boundary. The correspondence is a very
important method for strongly coupled field theories, and vice versa for
string theory and quantum gravity. Integrability \cite{Beisert:2010jr} has
greatly increased our understanding of the gauge-gravity correspondence. The
correspondence also reveals the nature of the emergent spacetimes e.g. \cite%
{Rangamani:2016dms}-\cite{Balasubramanian:2005mg}. The bulk emerges
dynamically from the quantum mechanical description that lives in fewer
dimensions. The correspondence unravels various emergent phenomena in
quantum gravitational theory and string theory, including the emergence of
the bulk spacetimes and excitations thereof.

On the gravity side, there are giant graviton branes \cite{McGreevy:2000cw}-%
\cite{Lin:2004nb}. These states in the Hilbert space of the quantum field
theory are explicitly mapped to the gravity side. There are branes in AdS
and in internal space. These giant gravitons are wrapped branes. They are
giant or large when compared to point gravitons. Analyses in the field
theory side show that these different states live in the same Hilbert space.
The dual large operators and their representation bases have been
illuminated \cite{Corley:2001zk}-\cite{Lin:2004nb}. These are brane-like
operators, which are holographically related to branes on the gravity side.
Moreover, these branes are related to emergent geometries on the bulk side.
Furthermore, branes in bulk side provide boundaries for strings and are
hence instrumental for analyzing boundary states and open strings, among
other things.

The countings of giant graviton brane states have been computed recently in
details \cite{Gaiotto:2021xce}-\cite{Liu:2022olj}, and show various precise
agreements with gravity considerations. On the other hand, direct
constructions of giant graviton operators have also been analyzed recently,
in details by \cite{Berenstein:2022srd}-\cite{Holguin:2022zii}. Various nice
ideas and techniques have been put forward, in order to make the
computations with large operators more efficient.

A new type of coherent states describing giant graviton branes were
constructed \cite{Berenstein:2022srd}. Various interesting aspects have been
analyzed, including, among other things \cite%
{Holguin:2022drf,Lin:2022wdr,Holguin:2022zii}, adding strings and
three-point functions. In particular, the correlation functions involving
the above coherent states describing the giant gravitons have been computed,
and agree with gravity exactly \cite{Holguin:2022zii}.

We generalize a construction of coherent state operators describing various
giant graviton branes. The set-ups in this paper help us to address the
questions how do physics of the bulk emerge from dual quantum theory on the
boundary. We describe the detailed construction in Sec. 2. We also add
strings and capture near-BPS states. The coherent states are BPS, hence we
can extract the near BPS spectrum subtracted from the BPS coherent state
backgrounds. In Sec. 3, we build two auxiliary field models, as convenient
intermediate steps to construct various operators and aim at unifying
different types of operators.

In Sec. 4.1 and 4.2, we package the coherent state parameters into
supermatrix. We also define new classes of coherent states. In Sec. 5, we
describe multi-matrix operators and BPS states, and unify some classes of
operators. Then in Sec. 6, we work on reduced coherent states as well as
their fermionic counterparts. Finally in Sec. 7, we make conclusions and
discuss some closely related aspects. More details of computations are
included in Appendices A to E.

\section{Fermionic counterparts of BPS coherent states}

\label{sec 2} \renewcommand{\theequation}{2.\arabic{equation}} %
\setcounter{equation}{0} \renewcommand{\thethm}{2.\arabic{thm}} %
\setcounter{thm}{0} \renewcommand{\theprop}{2.\arabic{prop}} %
\setcounter{prop}{0}

We generalize the construction of \cite{Berenstein:2022srd} by including
giants and dual giants simultaneously. We use, in addition to $N\times N$
coherent state parameter (CSP) $\Lambda $, also $T\times T$ coherent state
parameter$~\Lambda ^{c}$ recording the information of giants. $T$ is a
general integer, which has independence with respective to $N$. The state
has both $U(N)~$invariance and hidden $U(T)$ invariance. The state is a
class of states describing at most $N$ dual giants and at most $T$ giants
simultaneously. The $U(T)$ is a hidden symmetry that is the symmetry
enhanced by at most $T$ giants.

We consider the following operators
\begin{equation}
O=\int dUd\chi ^{\dagger }d\chi e^{\mathrm{tr}(\chi \chi ^{\dagger })}\exp (%
\mathrm{tr}(\alpha U\Lambda U^{\dagger }Z-\beta \chi \Lambda ^{c}\chi
^{\dagger }Z)).
\end{equation}%
$U$ is an auxiliary matrix which is $N\times N$ unitary matrix. The fermion
auxiliary field $\chi $ is $N\times T$. They are equivalent to $T$ column
vectors. We denote their conjugates as $\chi ^{\dagger }$ which is $T\times N
$.$~$The extra coherent state parameter $\Lambda ^{c}$ is $T\times T$.~$%
\Lambda ^{c}$ contains the information of internal giants. The trace is over
$N$, $\mathrm{tr}_{N}$.$~$The construction is for general $T,N$. Due to that
fermions are anti-commuting variables \cite{Berezin,Berezin:87}, $\mathrm{tr}%
_{N}\chi \chi ^{\dagger }=-\mathrm{tr}_{T}\chi ^{\dagger }\chi $, or $\chi
_{\alpha i}\chi _{i}^{\dagger \alpha }=-\chi _{i}^{\dagger \alpha }\chi
_{\alpha i}$. The fermion statistics are important here. In the above, $%
\alpha =1,...,N$, and $i=1,...,T$. Note%
\begin{equation}
\int d\chi ^{\dagger }d\chi e^{\mathrm{tr}(\chi \chi ^{\dagger })}=1,~\int
d\chi ^{\dagger }d\chi =0.
\end{equation}

Now we generalize to multi-matrix case or eighth BPS case,%
\begin{eqnarray}
&&O=\int dUd\chi ^{\dagger }d\chi e^{\mathrm{tr}(\chi \chi ^{\dagger })}\exp
(\mathrm{tr}(\alpha U\Lambda _{z}U^{\dagger }Z+\alpha U\Lambda
_{y}U^{\dagger }Y+\alpha U\Lambda _{x}U^{\dagger }X  \notag \\
&&-\beta \chi \Lambda _{z}^{c}\chi ^{\dagger }Z-\beta \chi \Lambda
_{y}^{c}\chi ^{\dagger }Y-\beta \chi \Lambda _{x}^{c}\chi ^{\dagger }X)).
\label{operators}
\end{eqnarray}%
The operators (\ref{operators}) are fermionic extension of the eighth BPS
coherent states. The operators are parametrized by six matrices. $\Lambda
_{z},\Lambda _{y},\Lambda _{x}~$are three mutually commuting $N\times N~$%
matrices. $\Lambda _{z}^{c},\Lambda _{y}^{c},\Lambda _{x}^{c}~$are three
mutually commuting $T\times T~$matrices. $\Lambda _{z},\Lambda _{y},\Lambda
_{x}$ records the information of dual giants and $\Lambda _{z}^{c},\Lambda
_{y}^{c},\Lambda _{x}^{c}~$records the information of giants. This is for
general $T,N$. The operators have a hidden $U(T)$ symmetry. For $U_{c}\in
U(T)$, $U_{c}$ is $T\times T~$here, the operators have symmetry under$~\chi
\rightarrow \chi U_{c}$, $\chi ^{\dagger }\rightarrow U_{c}^{\dagger }\chi
^{\dagger }$, $\Lambda _{z,y,x}^{c}\rightarrow U_{c}^{\dagger }\Lambda
_{z,y,x}^{c}U_{c}$.

For general matrix parameters, with mutually commuting matrices, they are
eighth-BPS in general \cite{Berenstein:2022srd,Lin:2022wdr}. We may set $%
\Lambda _{x},\Lambda _{x}^{c}$ to zero, so they also include quarter-BPS.

Although $\Lambda _{(z,y,x)}^{c}$ are $T\times T$, we can make different
choices of the ranks of $\Lambda _{(z,y,x)}^{c}$ and denote the rank of
them, i.e. $\mathrm{rk\ }\Lambda _{(z,y,x)}^{c}$, as $p_{21},p_{22},p_{23}$,
and these numbers are between 0 and $T$. States with different $\mathrm{rk~}%
\Lambda _{(z,y,x)}^{c}$ are different types of states, describing different
numbers of giants with different momenta. $\Lambda _{z,y,x}^{c}$ also have a
rich sub-block structure, and the underlying states with different
sub-blocks have invariance under subgroup symmetries. Since e.g. $%
p_{2i}\leqslant T$, there are at most $T$ giants. Big $T$ can be viewed as a
regulator. As $T\rightarrow \infty $, the states include all possible number
of internal giants and dual giants.

The operator itself is a coherent state \cite{Glauber:1963fi} operator. It
can also be used as a generating function, since it can be expanded as the
linear superpositions of other multi-trace operators.

For special case $T=N$, we can write it in a different way,
\begin{eqnarray}
&&O=\int dUd\chi ^{\dagger }d\chi e^{\mathrm{tr}(\chi \chi ^{\dagger })}\exp
(\mathrm{tr}(\alpha U\Lambda _{z}U^{\dagger }Z+\alpha U\Lambda
_{y}U^{\dagger }Y+\alpha U\Lambda _{x}U^{\dagger }X  \notag \\
&&+\beta \Lambda _{z}^{c}\chi ^{\dagger }Z\chi +\beta \Lambda _{y}^{c}\chi
^{\dagger }Y\chi +\beta \Lambda _{x}^{c}\chi ^{\dagger }X\chi )),~~~\ ~~~T=N.
\end{eqnarray}%
Note the sign in front of $\beta$ in the definition, due to anti-commuting
nature of fermions.

The un-integrated operators can be used to calculate correlation functions.
The integrands in (\ref{operators}) are unintegrated operators. One of the
advantages of this set-up is that the un-integrated state is coherent state,
which has nice behaviors. For simplicity, we first illustrate with the
purely fermionic state,
\begin{eqnarray}
O &=&\int d\chi ^{\dagger }d\chi e^{\mathrm{tr}(\chi \chi ^{\dagger })}\exp
(-\mathrm{tr}(\beta \chi \Lambda _{z}^{c}\chi ^{\dagger }Z+\beta \chi
\Lambda _{y}^{c}\chi ^{\dagger }Y+\beta \chi \Lambda _{x}^{c}\chi ^{\dagger
}X)) \\
&=&\int d\chi ^{\dagger }d\chi e^{\mathrm{tr}(\chi \chi ^{\dagger })}O_{0}~.
\end{eqnarray}%
$O_{0}$ are un-integrated operators. The two-point function of the
un-integrated operator is $\langle O_{0}|O_{0}\rangle $, which is easier to
compute than the integrated one. Define
\begin{equation}
G=\int d\chi ^{\prime \dagger }d\chi ^{\prime }e^{\mathrm{tr}(\chi ^{\prime
}\chi ^{\prime \dagger })}\langle O_{0}|O_{0}\rangle ,
\end{equation}%
hence by integration, the two-point function of the integrated operator is%
\begin{equation}
\langle O|O\rangle =\int d\chi ^{\dagger }d\chi e^{\mathrm{tr}(\chi \chi
^{\dagger })}G~.
\end{equation}%
We illustrate this idea or method in Appendix A.

The method can also be used to calculate higher point functions. For
example. $W$ collectively denotes other fields. We could add other words $W$
into the un-integrated operators $O_{0i}(\chi ,\chi ^{\dagger },Z,Y,X;W),$
\begin{equation}
O_{i}=\int d\chi ^{\dagger }d\chi e^{\mathrm{tr}(\chi \chi ^{\dagger
})}O_{0i}(\chi ,\chi ^{\dagger },Z,Y,X;W).
\end{equation}%
\begin{equation}
G=G(\chi ,\chi ^{\dagger })=\int d\chi ^{\prime \prime \dagger }d\chi
^{\prime \prime }e^{\mathrm{tr}(\chi ^{\prime \prime }\chi ^{\prime \prime
\dagger })}\int d\chi ^{\prime \dagger }d\chi ^{\prime }e^{\mathrm{tr}(\chi
^{\prime }\chi ^{\prime \dagger })}\langle O_{03}^{\dagger
}O_{02}O_{01}\rangle .
\end{equation}%
\begin{equation}
\langle O_{3}^{\dagger }O_{2}O_{1}\rangle =\int d\chi ^{\dagger }d\chi e^{%
\mathrm{tr}(\chi \chi ^{\dagger })}G(\chi ,\chi ^{\dagger }).
\end{equation}%
We illustrate this idea or method in Appendix A. One can calculate more
inner products and correlation functions, using this representation (\ref%
{operators}).

It is also convenient to add strings. Here\textbf{\ }$_{i}\left( U^{\dagger
}W_{k}U\right) _{j}$ is a string with two boundaries stretching between
brane $i$ and $j$. If we take a trace, then, \textrm{tr}$W_{k}$,~\textrm{tr}$%
W_{k}^{\prime }$ become a closed string. The auxiliary fields serve to cut
the trace open, e.g. \cite%
{Berenstein:2022srd,Gaiotto:2021xce,Jiang:2019xdz,Chen:2019gsb,Lin:2022wdr}.
For any $N$, $T$, one can add strings. The background state here is new. $%
U^{\dagger }W_{k}U$ and $\chi ^{\dagger }W_{k}^{\prime }\chi ~$play the role
of cutting the trace open. $k$ labels the $k$-th word. We can have
generating function for adding strings,%
\begin{eqnarray}
\tilde{F}[t_{k},s_{k}] &=&\int dUd\chi ^{\dagger }d\chi e^{\mathrm{tr}(\chi
\chi ^{\dagger })}\exp (\Gamma )\exp (\mathrm{tr}(\sum\limits_{k}Ut_{k}U^{%
\dagger }W_{k}-\sum\limits_{k}\chi s_{k}\chi ^{\dagger }W_{k}^{\prime })),
\notag \\
\Gamma &=&\mathrm{tr}(\alpha \Lambda _{z}U^{\dagger }ZU+\alpha \Lambda
_{y}U^{\dagger }YU+\alpha \Lambda _{x}U^{\dagger }XU-\beta \chi \Lambda
_{z}^{c}\chi ^{\dagger }Z  \notag \\
&&-\beta \chi \Lambda _{y}^{c}\chi ^{\dagger }Y-\beta \chi \Lambda
_{x}^{c}\chi ^{\dagger }X).  \label{adding strings}
\end{eqnarray}%
Here $s_{k}$ is $T\times T$ matrix and $t_{k}$ is $N\times N$ matrix. As a
side-remark, $\Gamma $ here resembles the role of $S_{eff}$. Taking
differentials,
\begin{equation}
\delta _{\left( t_{k}\right) _{j}^{i}}\tilde{F}[t_{k},s_{k}]=\int dUd\chi
^{\dagger }d\chi e^{\mathrm{tr}(\chi \chi ^{\dagger })}\exp (\Gamma
)~_{i}\left( U^{\dagger }W_{k}U\right) _{j},
\end{equation}%
\begin{equation}
\delta _{\left( s_{k}\right) _{j_{l}}^{i_{l}}}\cdot \cdot \cdot \delta
_{\left( t_{k}\right) _{j_{1}}^{i_{1}}}\tilde{F}[t_{k},s_{k}]=\int dUd\chi
^{\dagger }d\chi e^{\mathrm{tr}(\chi \chi ^{\dagger })}\exp (\Gamma
)~_{i_{1}}\left( U^{\dagger }W_{k}U\right) _{j_{1}}\cdot \cdot \cdot
~_{i_{l}}\left( \chi ^{\dagger }W_{k}^{\prime }\chi \right) _{j_{l}}.
\end{equation}%
Hence one can perform adding-string each time taking the differentials.

The string-added states capture near-BPS states. The coherent state is
itself BPS, hence we can extract the near BPS spectrum subtracted from the
BPS coherent state background. The excitation of the states are expected to
include giant magnons, spinning strings and other strings, as well as in
close connection to $SU(2|2)$ symmetry \cite%
{Beisert:2005tm,Berenstein:2014zxa,deCarvalho:2020pdp,Berenstein:2020grg,Gadde:2010ku}
and $SL(2)$ sectors and their cousins.

\section{Auxiliary field models for constructions of states}

\label{sec 3} \renewcommand{\theequation}{3.\arabic{equation}} %
\setcounter{equation}{0} \renewcommand{\thethm}{3.\arabic{thm}} %
\setcounter{thm}{0} \renewcommand{\theprop}{3.\arabic{prop}} %
\setcounter{prop}{0}

The operators in \cite{Berenstein:2022srd} and in (\ref{operators}) of Sec.
2 are in the form of the exponent of a trace. This trace on the exponent is
reminiscent of an effective action. Hence one may introduce auxiliary fields
coupled to the main fields appearing in the operator, and then integrate the
auxiliary fields to produce an effective action of the main fields. Here we
analyze two auxiliary field models, aiming to unify different types of
operators. The auxiliary field models serve as a intermediate way to
construct operators, and it can unify different types of operators. The
auxiliary fields here serve to cut the trace open as well as into two halves.

We first discuss the first auxiliary field model as follows. Here $\chi
,\psi ,\phi ,\varphi $ are complex fields. The $\phi ,\varphi ~$are bosons
and are$~N\times p_{1}~$matrices, they are composed of $p_{1}$~$N\times 1~$%
vectors. We denote the components of them, as $\phi _{i},\varphi
_{i},1\leqslant i\leqslant p_{1}$. The $\chi ,\psi ~$are fermions and are$%
~N\times p_{2}~$matrices, they are composed of $p_{2}$~$N\times 1~$vectors,
and we denote the components of them, as $\chi _{i},\psi _{i},1\leqslant
i\leqslant p_{2}$. The fermions $\chi ,\psi $ are anti-commuting variables,
e.g. \cite{Berezin,Berezin:87}. $\Lambda _{1i}$ are $p_{1}~N\times N~$%
matrices. $\Lambda _{2i}$ are $p_{2}~N\times N~$matrices. $U $ is $N\times N$
unitary matrix. For simplicity of the notation, Einstein summation
convention is used. The operator is
\begin{eqnarray}
&&\int d\chi ^{\dagger }d\chi d\psi ^{\dagger }d\psi d\phi ^{\dagger }d\phi
d\varphi ^{\dagger }d\varphi dU  \notag \\
&&\exp (\sum\limits_{i=1}^{p_{1}}-\varphi _{i}^{\dagger }\phi _{i}-\varphi
_{i}\phi _{i}^{\dagger }+\alpha \phi _{i}^{\dagger }ZU\phi _{i}+\varphi
_{i}^{\dagger }\Lambda _{1i}U^{\dagger }\varphi _{i}  \notag \\
&&+\sum\limits_{i=1}^{p_{2}}-\chi _{i}^{\dagger }\psi _{i}-\psi
_{i}^{\dagger }\chi _{i}+\beta \psi _{i}^{\dagger }ZU\psi _{i}+\chi
_{i}^{\dagger }\Lambda _{2i}U^{\dagger }\chi _{i}).  \label{operators aux}
\end{eqnarray}%
Due to anti-commuting nature of fermions, we have e.g. $\sum\limits_{i,%
\alpha }\phi _{i\alpha }\varphi _{i}^{\dagger \alpha }=\sum\limits_{i,\alpha
}\varphi _{i}^{\dagger \alpha }\phi _{i\alpha }$, while $\sum\limits_{i,%
\alpha }\psi _{i\alpha }\chi _{i}^{\dagger \alpha }=-\sum\limits_{i,\alpha
}\chi _{i}^{\dagger \alpha }\psi _{i\alpha }$.

Hence we define (\ref{operators aux}) as a state. If we integrate out $\chi
,\psi ,\phi ,\varphi $, we get the integrand inside the $U$-integral
\begin{equation}
\int dU\frac{{\prod_{i=1}^{p_{2}}}\det \left[ 1-\beta U\Lambda
_{2i}U^{\dagger }Z\right] }{{\prod_{i=1}^{p_{1}}}\det \left[ 1-\alpha
U\Lambda _{1i}U^{\dagger }Z\right] }.  \label{operators 04}
\end{equation}%
The determinant in the denominator can be expanded and describe dual giants.

Now we consider simpler cases. Consider $\Lambda _{1i}$ are equal to $%
\Lambda _{1}$, and $\Lambda _{2i}$ are equal to $\Lambda _{2}$, the states
are
\begin{equation}
\int dU\frac{\det \left[ 1-\beta U\Lambda _{2}U^{\dagger }Z\right] ^{p_{2}}}{%
\det \left[ 1-\alpha U\Lambda _{1}U^{\dagger }Z\right] ^{p_{1}}}.
\end{equation}%
Consider $\Lambda _{1}\alpha ,\beta \Lambda _{2}$ are expansion parameters.
We can expand the states using traces. Define new parameter matrix $\alpha
\Lambda :=\alpha p_{1}\Lambda _{1}-\beta p_{2}\Lambda _{2}$. We have the
universal limit \cite{Berenstein:2022srd}
\begin{equation}
\int dU\exp (\mathrm{tr}(\alpha U\Lambda U^{\dagger }Z).
\label{operators 05}
\end{equation}

Then consider $\Lambda _{1i}=\lambda _{+i}I,\Lambda _{2i}=\lambda _{-i}I$.
The integrand is
\begin{equation}
\frac{{\prod_{i=1}^{p_{2}}}\det \left[ 1-\beta Z\lambda _{-i}\right] }{{%
\prod_{i=1}^{p_{1}}}\det \left[ 1-\alpha Z\lambda _{+i}\right] }=\mathcal{N}%
\frac{{\prod_{i=1}^{p_{2}}}\det \left[ Z-u_{i}\right] }{{\prod_{i=1}^{p_{1}}}%
\det \left[ Z-\xi _{i}\right] }  \label{operators 06}
\end{equation}%
with%
\begin{equation}
u_{i}=\lambda _{-i}^{-1}\beta ^{-1},~~\xi _{i}=\lambda _{+i}^{-1}\alpha
^{-1}.
\end{equation}%
Note that the determinant in the denominator may be expanded out as, e.g.%
\begin{equation}
\det {}^{-1}[Z-\xi ]=(-\xi )^{-N}\sum\limits_{n=0}^{\infty }\xi ^{-n}\chi
_{(n)}(Z),
\end{equation}%
where $\chi _{(n)}$ is Schur polynomial of fully symmetric representation,
and when $\xi \rightarrow \infty $ the operator reduces to identity
operator. As a side-remark, when $p_{1},p_{2\text{ }}$are equal, (\ref%
{operators 06}) may be written in terms of super-determinants, see also
discussion in related context in Appendix D.

As described above, the states (\ref{operators 04}) have two nice limits.
One limit interpolates the BPS coherent state operators (\ref{operators 05}%
). Another limit interpolates the multi-determinantal operators or states (%
\ref{operators 06}).

The auxiliary fields in (\ref{operators aux}) may seem doubled and can be
shortened to half. This is because we need the extra half of them to cut the
trace open and into two halves. Alternatively, we may integrate out half of
the total auxiliary fields. Now redefine linear combinations\ $\tilde{\psi}=%
\frac{1}{\sqrt{2}}(\chi +\psi )$,$~\tilde{\phi}=\frac{1}{\sqrt{2}}(\varphi
+\phi )$. Integrate out the other two linear combinations $\frac{1}{\sqrt{2}}%
(\chi -\psi )$,$\frac{1}{\sqrt{2}}(\varphi -\phi )$. And then we simplify
the symbols by removing tilde symbols, to denote $\tilde{\psi}$ as $\psi $,
and denote $\tilde{\phi}$ as $\phi $. $\sum\limits_{i=1}^{p_{1}},\sum%
\limits_{i=1}^{p_{2}}$ is equivalent to tr$_{p_{1}},$tr$_{p_{2}}$. Without
inserting strings, the state is
\begin{equation}
\int d\phi ^{\dagger }d\phi d\psi ^{\dagger }d\psi dU\exp
\sum\limits_{i=1}^{p_{1}}(-\phi _{i}^{\dagger }\phi _{i}+\alpha \phi
_{i}^{\dagger }\Lambda _{1,i}U^{\dagger }ZU\phi
_{i})+\sum\limits_{i=1}^{p_{2}}(-\psi _{i}^{\dagger }\psi _{i}+\beta \psi
_{i}^{\dagger }\Lambda _{2,i}U^{\dagger }ZU\psi _{i}).
\end{equation}

We can adding strings to giants and dual giants. We can first add strings
before doing the $U$-integral, and then in a later stage do the $U$
integral. Let's denote by $\left\vert v\right\rangle _{a},\left\vert
v\right\rangle _{b}$ the eigenvectors corresponding to $a,b$ eigenvalues
\cite{Berenstein:2022srd,Lin:2022wdr}. We add the strings corresponding to
words $W_{k}$. We use the following notation, written in component-form:
\begin{equation}
_{a_{1}}\left( \phi _{i_{1}}^{\dagger a_{1}}U^{\dagger }W_{1}U\phi
_{i_{1}b_{1}}\right) _{b_{1}}.  \label{words 02}
\end{equation}%
The state with strings added is
\begin{eqnarray}
&&\int d\phi ^{\dagger }d\phi d\psi ^{\dagger }d\psi dU  \notag \\
&&\exp (\sum\limits_{i=1}^{p_{1}}(-\phi _{i}^{\dagger }\phi _{i}+\alpha \phi
_{i}^{\dagger }\Lambda _{1,i}U^{\dagger }ZU\phi
_{i})+\sum\limits_{i=1}^{p_{2}}(-\psi _{i}^{\dagger }\psi _{i}+\beta \psi
_{i}^{\dagger }\Lambda _{2,i}U^{\dagger }ZU\psi _{i}))~  \notag \\
&&\left( \phi _{i_{1}}^{\dagger a_{1}}U^{\dagger }W_{1}U\phi
_{i_{1}b_{1}}\right) \cdot \cdot \cdot \left( \psi _{i_{l}}^{\dagger
a_{l}}U^{\dagger }W_{l}U\psi _{i_{l}b_{l}}\right) .
\end{eqnarray}%
We can generalize them to quarter BPS and eighth BPS cases, e.g.
\begin{eqnarray}
&&\int d\phi ^{\dagger }d\phi d\psi ^{\dagger }d\psi dU  \notag \\
&&\exp (\sum\limits_{i=1}^{p_{1}}(-\phi _{i}^{\dagger }\phi _{i}+\alpha \phi
_{i}^{\dagger }\left( \Lambda _{z1,i}U^{\dagger }ZU+\Lambda
_{y1,i}U^{\dagger }YU\right) \phi _{i})  \notag \\
&&+\sum\limits_{i=1}^{p_{2}}(-\psi _{i}^{\dagger }\psi _{i}+\beta \psi
_{i}^{\dagger }\left( \Lambda _{z2,i}U^{\dagger }ZU+\Lambda
_{y2,i}U^{\dagger }YU\right) \psi _{i}))~  \notag \\
&&\left( \phi _{i_{1}}^{\dagger a_{1}}U^{\dagger }W_{1}U\phi
_{i_{1}b_{1}}\right) \cdot \cdot \cdot \left( \psi _{i_{l}}^{\dagger
a_{l}}U^{\dagger }W_{l}U\psi _{i_{l}b_{l}}\right) .
\end{eqnarray}

We now discuss a second auxiliary field model with complementary coherent
state parameter matrix. $\Lambda _{2i}^{c}$ are $p\times p$ matrices, with $%
p=p_{2}$. $\Lambda _{2}^{c}$ is complementary to $\Lambda _{1}$, and the
label $c$ refers to complementary, because it's the fermionic counterparts
to $\Lambda _{1}$. The$\ \Lambda _{2}^{c}$ plays the role of $\Lambda _{2}$,
but in an alternative formulation; it plays the same role of introducing
internal giants. $\Lambda _{2}^{c}$ is $p\times p$ while $\Lambda _{2}$ is $%
N\times N$. The integrand is
\begin{equation}
\exp (\sum\limits_{i=1}^{p_{1}}(-\phi _{i}^{\dagger }\phi _{i}+\alpha \phi
_{i}^{\dagger }\Lambda _{1,i}U^{\dagger }ZU\phi
_{i})+\sum\limits_{i=1}^{p_{2}}(-\psi _{i}^{\dagger }\psi _{i}+\beta \Lambda
_{2,i}^{c}\psi _{i}^{\dagger }Z\psi _{i})).
\end{equation}

We can add strings. We use the following notation, written in component
form. In addition to the words in (\ref{words 02}), we also add the words:
\begin{equation}
_{a_{2}}\left( \psi _{j_{1}}^{\dagger a_{2}}W_{2}^{\prime }\psi
_{j_{1}b_{2}}\right) _{b_{2}}.
\end{equation}%
The state with strings added is%
\begin{eqnarray}
&&\int d\phi d\phi ^{\dagger }d\psi d\psi ^{\dagger }dU  \notag \\
&&\exp (\sum\limits_{i=1}^{p_{1}}(-\phi _{i}^{\dagger }\phi _{i}+\alpha \phi
_{i}^{\dagger }\Lambda _{1,i}U^{\dagger }ZU\phi
_{i})+\sum\limits_{i=1}^{p_{2}}(-\psi _{i}^{\dagger }\psi _{i}+\beta \Lambda
_{2,i}^{c}\psi _{i}^{\dagger }Z\psi _{i}))~  \notag \\
&&\left( \phi _{i_{1}}^{\dagger a_{1}}U^{\dagger }W_{1}U\phi
_{i_{1}b_{1}}\right) \left( \psi _{j_{1}}^{\dagger a_{2}}W_{2}^{\prime }\psi
_{j_{1}b_{2}}\right) \cdot \cdot \cdot
\end{eqnarray}%
One can have multiple words, $W_{3},W_{4}$, ... and so on \cite%
{Berenstein:2022srd,Lin:2022wdr}. One can also generalize them to the
quarter BPS and eighth BPS case by global symmetry considerations on
matrices $Z,Y,X$ and we will discuss them more in Sec. 5.

\section{HCIZ super integral related states}

\label{sec 4} \renewcommand{\theequation}{4.\arabic{equation}} %
\setcounter{equation}{0} \renewcommand{\thethm}{4.\arabic{thm}} %
\setcounter{thm}{0} \renewcommand{\theprop}{4.\arabic{prop}} %
\setcounter{prop}{0}

\subsection{General $N,T$ cases}

The complementary coherent state parameters $\Lambda^{c}$ are $T\times T$
matrices. Moreover, there is expectation that there is emergent or enhanced
symmetry of $T$ giants, with $U(T)$ hidden symmetry, which is also evident
from the properties of their near-BPS excitations.

We use hidden $U(N)\times U(T)$ symmetry to describe $T$ giants and $N$ dual
giants. We have included the situations that the coherent state parameters
are not maximal-rank, like the ones discussed in Sec. 2. In these cases, we
can have e.g. sub-symmetry $U(p_{1})\times U(p_{2})\subseteq U(N)\times U(T)$%
. There is a rich sub-block structure of $\Lambda ^{c}$, and when breaking
to subsymmetry, ${\prod_{i=1}^{n}}U(p_{2i})\subseteq U(T)$.

We arrange the coherent state parameter as supermatrix,
\begin{equation}
\Lambda _{S}=\left(
\begin{array}{cc}
\alpha \Lambda & 0 \\
0 & -\beta \Lambda ^{c}%
\end{array}%
\right) .
\end{equation}%
The subscript $S$ refers to supermatrix. $\Lambda $ is $N\times N$ and $%
\Lambda ^{c}$ is $T\times T$.

We could consider $Z_{S}=\left(
\begin{array}{cc}
Z & 0 \\
0 & \Xi%
\end{array}%
\right) $, where $\Xi$ is a constant auxiliary field matrix, and we put it
as constant since its role is to shift auxiliary fermion mass terms by the
amount $\alpha \mathrm{tr}_{T}(\Xi \chi ^{\dagger }\Lambda \chi )$ in the
exponent, as well as an overal constant scale factor unrelated to other
fields. For simplicity we set $\Xi =0$. Hence we consider
\begin{equation}
Z_{S}=\left(
\begin{array}{cc}
Z & 0 \\
0 & 0%
\end{array}%
\right) .
\end{equation}

The $\left( N+T\right) \times \left( N+T\right) ~$super unitary matrix can
be represented by $U_{S}=\exp (iH_{S})$, where $H_{S}$ is a $\left(
N+T\right) \times \left( N+T\right) ~$hermitian super matrix. With our
convention, $U_{S}=\left(
\begin{array}{cc}
U & if \\
if^{\dagger } & U_{c}%
\end{array}%
\right) $, and we later identify$~f^{\dagger }=\chi ^{\dagger }$. Here $f$
is $N\times T$.

We draw inspiration from HCIZ integral for superunitary matrices. Consider
the following integral
\begin{equation}
I=\ \int [dU_{S}]\exp (\mathrm{Str}(U_{S}\Lambda _{S}U_{S}^{\dagger }Z_{S})).
\end{equation}%
We can use this integrand to define coherent states of the supermatrix $%
Z_{S} $, which is in general $\left( N+T\right) \times \left( N+T\right) $
matrices. The integrand is $\exp (\mathrm{Str}(\Lambda _{S}U_{S}^{\dagger
}Z_{S}U_{S}))$ = $\exp (\alpha \mathrm{tr}_{N}(\Lambda U^{\dagger }ZU)~$+$%
~\beta \mathrm{tr}_{T}(\Lambda ^{c}\chi ^{\dagger }Z\chi ))$.

This leads us to define the coherent states as follows. The quantities
inside the square brackets denote the four diagonal blocks of the
supermatrices,
\begin{eqnarray}
&&O_{1}[\alpha \Lambda ,-\beta \Lambda ^{c};Z,0]  \notag \\
&=&\int dUd\chi ^{\dagger }d\chi e^{\mathrm{tr}(\chi \chi ^{\dagger })}\exp (%
\mathrm{tr}_{N}(\alpha U\Lambda U^{\dagger }Z-\beta \chi \Lambda ^{c}\chi
^{\dagger }Z)).
\end{eqnarray}%
$\alpha ,\beta $ are real, and play the role of inverse temperature, or
fugacity. We keep them to keep track of which field is which. In principle, $%
\alpha ,\beta $ can be scaled away by redefinitions, e.g. from rescaling of
coherent state parameters. Combining the ideas and methods in \cite%
{Berenstein:2022srd}, we can make use of HCIZ-type super integral to
calculate inner products and correlation functions.

We can generalize the states to the cases with more supermatrices. The
following $\Lambda _{z,y,x}^{S},Z_{S},Y_{S},X_{S}$ are $\left( N+T\right)
\times \left( N+T\right) $ supermatrices.%
\begin{equation}
\Lambda _{z}^{S}=\left(
\begin{array}{cc}
\alpha \Lambda _{z} & 0 \\
0 & -\beta \Lambda _{z}^{c}%
\end{array}%
\right) ,\Lambda _{y}^{S}=\left(
\begin{array}{cc}
\alpha \Lambda _{y} & 0 \\
0 & -\beta \Lambda _{y}^{c}%
\end{array}%
\right) ,\Lambda _{y}^{S}=\left(
\begin{array}{cc}
\alpha \Lambda _{x} & 0 \\
0 & -\beta \Lambda _{x}^{c}%
\end{array}%
\right) .  \label{op 13}
\end{equation}%
\begin{eqnarray}
~~Z_{S} &=&\left(
\begin{array}{cc}
Z & 0 \\
0 & 0%
\end{array}%
\right) ,Y_{S}=\left(
\begin{array}{cc}
Y & 0 \\
0 & 0%
\end{array}%
\right) ,~X_{S}=\left(
\begin{array}{cc}
X & 0 \\
0 & 0%
\end{array}%
\right) .  \label{op 15} \\
I[\Lambda _{z,y,x}^{S}] &=&\int [dU_{S}]\exp (\mathrm{Str}(\Lambda
_{z}^{S}U_{S}^{\dagger }Z_{S}U_{S}+\Lambda _{z}^{S}U_{S}^{\dagger
}Y_{S}U_{S}+\Lambda _{z}^{S}U_{S}^{\dagger }X_{S}U_{S})).
\end{eqnarray}%
Note that it is the same $U_{S}$ that couples to the three fields. In
component form, the operators are
\begin{eqnarray}
&&O_{1}[\alpha \Lambda _{z,y,x},-\beta \Lambda _{z,y,x}^{c};Z,Y,X;0]  \notag
\\
&=&\int dUd\chi ^{\dagger }d\chi e^{\mathrm{tr}(\chi \chi ^{\dagger })}\exp (%
\mathrm{tr}_{N}(\alpha U\Lambda _{z}U^{\dagger }Z-\beta \chi \Lambda
_{z}^{c}\chi ^{\dagger }Z+\alpha U\Lambda _{y}U^{\dagger }Y-\beta \chi
\Lambda _{y}^{c}\chi ^{\dagger }Y  \notag \\
&&+\alpha U\Lambda _{x}U^{\dagger }X-\beta \chi \Lambda _{x}^{c}\chi
^{\dagger }X)).~~~~~
\end{eqnarray}%
Here there are six matrix parameters. Under parity-transposition of the
supermatrix, the two types of terms get swapped as $\Lambda
_{z,y,x}\leftrightarrow \Lambda _{z,y,x}^{c}$. For $N=T$ case, in Appendix B
we discuss a generalization by adding $\bar{Z},{\bar{Y},\bar{X}}$ in the
lower-right blocks.

There is a parity-transposition operation. It's a symmetry operation as
follows%
\begin{equation}
\Lambda \longleftrightarrow \Lambda ^{c},
\end{equation}%
which can switch giants and dual giants. We discuss more relation to this
operation in a more general setting in Appendix D. In the supermatrix
proposal, it's
\begin{equation}
BB\longleftrightarrow FF,~~BF\longleftrightarrow FB,
\end{equation}%
hence it's the switch of BB/FF in supermatrix. This is the
parity-transposition in the supermatrix. This adds some new perspectives for
helping to understand the giant-dual-giant duality-transformation.

\subsection{$N=T$ special cases}

We construct another type of states, for $N=T$ special cases,
\begin{eqnarray}
\Lambda _{S} &=&\left(
\begin{array}{cc}
\alpha \Lambda & 0 \\
0 & -\beta \Lambda ^{c}%
\end{array}%
\right) ,Z_{S}=\left(
\begin{array}{cc}
Z & 0 \\
0 & Y%
\end{array}%
\right) . \\
I &=&\int [dU_{S}]\exp (\mathrm{Str}(\Lambda _{S}U_{S}^{\dagger
}Z_{S}U_{S})).
\end{eqnarray}%
$Y$ denotes another general $N\times N$ matrix. For example, it can be
another scalar in the $SU(2)$ sector of $N=4$ YM. It can be other matrix
fields, e.g. $X,\bar{Z}$, and so on. $U_{c}$ is related to induced gauge
fields on internal giants. We denote the operators constructed by these
supermatrices as $O_{2}[\alpha \Lambda ,-\beta \Lambda ^{c};Z,Y],$
\begin{eqnarray}
&&O_{2}[\alpha \Lambda ,-\beta \Lambda ^{c};Z,Y]  \notag \\
&=&\int dU_{c}dUd\chi ^{\dagger }d\chi e^{\mathrm{tr}(\chi \chi ^{\dagger
})}\exp (\mathrm{tr}_{N}(\alpha U\Lambda U^{\dagger }Z-\beta \chi \Lambda
^{c}\chi ^{\dagger }Z)+\mathrm{tr}_{N}(\beta U_{c}\Lambda ^{c}U_{c}^{\dagger
}Y+\alpha \chi ^{\dagger }\Lambda \chi Y)).  \notag \\
&&  \label{operators 08}
\end{eqnarray}

$O_{2}[\alpha \Lambda ,-\beta \Lambda ^{c};Z,Y]$ contains four independent
matrices. Apart from the parity transposition operations described above,
they have some relations to the operators in Sec. 4.1.

For $Y$ being another complex scalar field in $N$=$4$ SYM, $O_{2}[\alpha
\Lambda ,-\beta \Lambda ^{c};Z,Y]$ is quarter BPS at free theory level, and
is also in the $SU(2)$ sector. Moreover, the state $O_{2}[\alpha \Lambda
,-\beta \Lambda ^{c};Z,0]$ is a limit of $O_{2}[\alpha \Lambda ,-\beta
\Lambda ^{c};Z,Y]~$when $Y\rightarrow 0$. This is the limit, when quarter
BPS goes to half BPS. The $O_{2}[\alpha \Lambda ,-\beta \Lambda ^{c};Z,Y]$
can also be generalized to eighth BPS, by permuting $Z,Y,X$ and
incorporating $U(3)$ global symmetry.

There are relations between $O_{1}$ in Sec. 4.1 and $O_{2}$ in this
subsection. The eighth BPS state in Sec. 4.1 is%
\begin{equation}
O_{1}[\alpha \Lambda _{z,y,x},-\beta \Lambda _{z,y,x}^{c};Z,Y,X].
\end{equation}%
The quarter BPS state in Sec. 4.1 is
\begin{equation}
O_{1}[\alpha \Lambda _{z,y},-\beta \Lambda _{z,y}^{c};Z,Y].
\end{equation}%
It's different from the free theory quarter BPS (\ref{operators 08}) in this
subsection. $O_{2}[\alpha \Lambda ,-\beta \Lambda ^{c};Z,Y]$ is related to
one-term super HCIZ and the quarter BPS $O_{1}[\alpha \Lambda _{z,y},-\beta
\Lambda _{z,y}^{c};Z,Y]$ is related to two-term HCIZ. They don't subsume
each other. They have some common special cases,%
\begin{equation}
O_{2}[\alpha \Lambda ,0;Z,Y]~~\sim ~~O_{1}[\alpha \Lambda _{z,y}=(\alpha
\Lambda ,0),-\beta \Lambda _{z,y}^{c}=(0,-\alpha \Lambda );Z,Y].
\end{equation}%
Note that as in \cite{Berenstein:2022srd}, quarter and eighth BPS coherent
states can be constructed by two-term and three-term HCIZs.

Now consider $O_{2}[\alpha \Lambda ,-\beta \Lambda ^{c};Z,Z]$ and this is
half BPS. This is the $T=N$ case.%
\begin{eqnarray}
&&O_{2}[\alpha \Lambda ,-\beta \Lambda ^{c};Z,Z]  \notag \\
&=&\int dU_{c}dUd\chi ^{\dagger }d\chi e^{\mathrm{tr}(\chi \chi ^{\dagger
})}\exp (\mathrm{tr}(\alpha U\Lambda U^{\dagger }Z-\beta \chi \Lambda
^{c}\chi ^{\dagger }Z+\beta U_{c}\Lambda ^{c}U_{c}^{\dagger }Z+\alpha \chi
^{\dagger }\Lambda \chi Z)).  \notag \\
&&
\end{eqnarray}%
In Appendix C, we discuss more detailed properties of this state. These
states are suggestive of underlying large $N$ eigenvalue density approach.

One can also add strings in the supermatrix cases. $W_{S}$ is a word, built
from linear combination of hermitian supermatrices. They include other
fields in the dual gauge theory. One may have a generating function $\tilde{F%
}[\tau _{S}]$, including insertion of $\tau _{S}U_{S}^{\dagger }W_{S}U_{S}$
at the exponent,%
\begin{equation}
\delta _{\tau _{S}}...\delta _{\tau _{S}}\tilde{F}[\tau _{S}]\sim \int
[dU_{S}]\exp (\mathrm{Str}(\Lambda _{S}U_{S}^{\dagger
}Z_{S}U_{S}))(U_{S}^{\dagger }W_{S}U_{S})~...~(U_{S}^{\dagger }W_{S}U_{S}).
\end{equation}%
This is similar to the component-forms in e.g. (\ref{adding strings}).

One finally also make $Z$ dynamical with its own action and couplings when
computing correlation functions with other extra operators, or spectrum of
states when adding extra words or strings. In other words, we do computation
of these using un-integrated operators, and then integrate the answer by the
integration of auxiliary variables. The $U$-integral may also be viewed as
this type of auxiliary variable integral. Here we also illustrate
example-calculation here in Appendix C using $O_{2}$.

\section{Multi-matrix states and BPS operators}

\label{sec 5} \renewcommand{\theequation}{5.\arabic{equation}} %
\setcounter{equation}{0} \renewcommand{\thethm}{5.\arabic{thm}} %
\setcounter{thm}{0} \renewcommand{\theprop}{5.\arabic{prop}} %
\setcounter{prop}{0}

We can make generalization to quarter BPS and eighth BPS states, from the
auxiliary field models. We add $U(2)$ global symmetry to the auxiliary model
in Sec. 3. From the first auxiliary model,%
\begin{eqnarray}
&&\int d\phi ^{\dagger }d\phi d\psi ^{\dagger }d\psi dU  \notag \\
&&\exp (\sum\limits_{i=1}^{p_{1}}(-\phi _{i}^{\dagger }\phi _{i}+\alpha \phi
_{i}^{\dagger }\left( \Lambda _{z1i}U^{\dagger }ZU+\Lambda _{y1i}U^{\dagger
}YU\right) \phi _{i})  \notag \\
&&+\sum\limits_{i=1}^{p_{2}}(-\psi _{i}^{\dagger }\psi _{i}+\beta \psi
_{i}^{\dagger }\left( \Lambda _{z2i}U^{\dagger }ZU+\Lambda _{y2i}U^{\dagger
}YU\right) \psi _{i})).
\end{eqnarray}%
The $U$-integral gives the operators%
\begin{equation}
\int dU\frac{{\prod_{i=1}^{p_{2}}}\det \left[ 1-\beta U\Lambda
_{z2i}U^{\dagger }Z-\beta U\Lambda _{y2i}U^{\dagger }Y\right] }{{%
\prod_{i=1}^{p_{1}}}\det \left[ 1-\alpha U\Lambda _{z1i}U^{\dagger }Z-\alpha
U\Lambda _{y1i}U^{\dagger }Y\right] }.  \label{operators 09}
\end{equation}%
The integrand is a matrix polynomial function or a matrix rational function.
Note that if $p_{1}=p_{2}=0$, the operator reduces to identity operator. The
determinant in the denominator may be expanded using $\det^{-1}(1+M)=\exp (-%
\mathrm{tr}\ln (1+M))$, or e.g. as in Sec. 3.

The states in \cite{Berenstein:2022srd} is the universal limit of states (%
\ref{operators 09}):%
\begin{equation}
\int dU\exp (\alpha \mathrm{tr}(U\Lambda _{z}U^{\dagger }Z)+\beta \mathrm{tr}%
(U\Lambda _{y}U^{\dagger }Y)),
\end{equation}%
with $\alpha \Lambda _{(z,y)}=\alpha \sum\limits_{i=1}^{p_{1}}\Lambda
_{(z,y)1i}-\beta \sum\limits_{i=1}^{p_{2}}\Lambda _{(z,y)2i}$. The states (%
\ref{operators 09}) can always produce the states in \cite%
{Berenstein:2022srd}, when expanded, hence the state in \cite%
{Berenstein:2022srd} is universal.

Take $\Lambda _{z1i}=\lambda _{z+i}I,\Lambda _{z2i}=\lambda _{z-i}I,\Lambda
_{y1i}=\lambda _{y+i}I,\Lambda _{y2i}=\lambda _{y-i}I$. Then the $U$%
-integral is simple and the operators are
\begin{equation}
\frac{{\prod_{i=1}^{p_{2}}}\det \left[ \lambda _{z-i}Z+\lambda _{y-i}Y-\beta
^{-1}\right] }{{\prod_{i=1}^{p_{1}}}\det \left[ \lambda _{z+i}Z+\lambda
_{y+i}Y-\alpha ^{-1}\right] }.
\end{equation}

Now we look at the BPS condition for quarter BPS case. We have the
correspondence $Z,Y\leftrightarrow a_{Z}^{\dagger },a_{Y}^{\dagger }$, $%
\partial _{Z},\partial _{Y}\leftrightarrow a_{Z},a_{Y}$. The action of $%
\left[ a_{Z},a_{Y}\right] $, or $\left[ \partial _{Z},\partial _{Y}\right] ~$%
on the state, is a sum of $p_{1}+p_{2}$ pieces. Each piece for one-loop
dilatation operator, always contains an overal factor $\beta ^{2}\left[
\Lambda _{z2i},\Lambda _{y2i}\right] $, $\alpha ^{2}\left[ \Lambda
_{z1i},\Lambda _{y1i}\right] $, $\alpha \beta \left[ \Lambda _{z1i},\Lambda
_{y2i}\right] $, $\alpha \beta \left[ \Lambda _{z2i},\Lambda _{y1i}\right] $
etc. For non-zero $\alpha ,\beta $, the sufficient condition for BPS is
\begin{equation}
\left[ \Lambda _{z1i},\Lambda _{y1i}\right] =\left[ \Lambda _{z2i},\Lambda
_{y2i}\right] =\left[ \Lambda _{z1i},\Lambda _{y2i}\right] =\left[ \Lambda
_{z2i},\Lambda _{y1i}\right] =0.
\end{equation}%
Similar analysis of the one-loop dilatation operator also works for the
three-matrix case. There are non-renormalization theorems in the $U(2)$
sector and in the $U(3)$ sector analyzed in details in \cite%
{Pasukonis:2010rv,Lewis-Brown:2020nmg}. The global symmetry is $U(2)$ in the
quarter BPS case and is $U(3)$ in the eighth BPS case. It had been proved in
\cite{Pasukonis:2010rv,Lewis-Brown:2020nmg} that, if the operators are
global-symmetry symmetrized, then the operators in the kernel of the
one-loop dilatation operator \cite{Beisert:2003tq} are also in the kernel of
the higher loop dilatation operators. In the universal limit, the sufficient
condition simplifies to \cite{Berenstein:2022srd}%
\begin{equation}
\left[ \Lambda _{z},\Lambda _{y}\right] =0.
\end{equation}

For the $U(3)$ global symmetry sector including $Z,Y,X$, it works in the
same way. These are explored in \cite{Berenstein:2022srd,Lin:2022wdr}. From
the first auxiliary model in Sec. 3, The eighth BPS case is:%
\begin{equation}
\int dU\frac{{\prod_{i=1}^{p_{2}}}\det \left[ 1-\beta U\Lambda
_{z2i}U^{\dagger }Z-\beta U\Lambda _{y2i}U^{\dagger }Y-\beta U\Lambda
_{x2i}U^{\dagger }X\right] }{{\prod_{i=1}^{p_{1}}}\det \left[ 1-\alpha
U\Lambda _{z1i}U^{\dagger }Z-\alpha U\Lambda _{y1i}U^{\dagger }Y-\alpha
U\Lambda _{x1i}U^{\dagger }X\right] }.  \label{operators 10}
\end{equation}%
The universal limit of these states (\ref{operators 10}) is \cite%
{Berenstein:2022srd}:%
\begin{equation}
\int dU\exp (\alpha \mathrm{tr}(U\Lambda _{z}U^{\dagger }Z)+\alpha \mathrm{tr%
}(U\Lambda _{y}U^{\dagger }Y)+\alpha \mathrm{tr}(U\Lambda _{x}U^{\dagger
}X)),  \label{operators 12}
\end{equation}%
with $\alpha \Lambda _{(z,y,x)}=\alpha \sum\limits_{i=1}^{p_{1}}\Lambda
_{(z,y,x)1i}-\beta \sum\limits_{i=1}^{p_{2}}\Lambda _{(z,y,x)2i}$. If we
take $\Lambda _{x1i},\Lambda _{x2i}=0$, (\ref{operators 10}) reduces to (\ref%
{operators 09}). On the other hand, if we take $\Lambda _{(z,y,x)1i}=\lambda
_{(z,y,x)+i}I,\Lambda _{(z,y,x)2i}=\lambda _{(z,y,x)-i}I$, the operators are
\begin{equation}
\frac{{\prod_{i=1}^{p_{2}}}\det \left[ \lambda _{z-i}Z+\lambda
_{y-i}Y+\lambda _{x-i}X-\beta ^{-1}\right] }{{\prod_{i=1}^{p_{1}}}\det \left[
\lambda _{z+i}Z+\lambda _{y+i}Y+\lambda _{x+i}X-\alpha ^{-1}\right] }.
\end{equation}

A interesting situation is
\begin{equation}
\int dU{\prod_{i=1}^{p}}\det \left[ 1-\beta U\Lambda _{zi}U^{\dagger
}Z-\beta U\Lambda _{yi}U^{\dagger }Y-\beta U\Lambda _{xi}U^{\dagger }X\right]
.
\end{equation}%
These states interpolate both the coherent states in \cite%
{Berenstein:2022srd} when expanded using $\det (1+M)=\exp \mathrm{tr}\ln
(1+M)\overset{\lim }{\rightarrow }\exp \mathrm{tr}M\mathrm{~}$and also the
multiple determinant states, e.g. \cite%
{Chen:2019gsb,Jiang:2019xdz,Gaiotto:2021xce,Corley:2001zk} when the $U$
integral becomes trivial, e.g. $\Lambda _{(z,y,x)i}=\lambda _{(z,y,x)i}I$.
They automatically include the cases when some of the $\lambda _{i}$ go to
zero, e.g. $p_{21}+p_{22}+p_{23}=p$, and we break to subsymmetry. We can
also consider the states
\begin{equation}
\int dU\prod_{i=1}^{p_{21}}\det \left[ 1-\beta ZU\Lambda _{zi}U^{\dagger }%
\right] \prod_{i=1}^{p_{22}}\det \left[ 1-\beta YU\Lambda _{yi}U^{\dagger }%
\right] \prod_{i=1}^{p_{23}}\det \left[ 1-\beta XU\Lambda _{xi}U^{\dagger }%
\right] .
\end{equation}%
The universal limit is (\ref{operators 12}) with $\alpha \Lambda
_{(z,y,x)}=-\beta \sum\limits_{i=1}^{p_{2(1,2,3)}}\Lambda _{(z,y,x)2i}$. If $%
\Lambda _{(z,y,x)2i}=\lambda _{(z,y,x)i}I$, the states are
\begin{equation}
\prod_{i=1}^{p_{21}}\det \left[ \lambda _{zi}Z-\beta ^{-1}\right]
\prod_{i=1}^{p_{22}}\det \left[ \lambda _{yi}Y-\beta ^{-1}\right]
\prod_{i=1}^{p_{23}}\det \left[ \lambda _{xi}X-\beta ^{-1}\right] .
\end{equation}%
They correspond to various states of giant gravitons with different numbers
and momenta.

Below, we also make other remarks in other perspectives. The operators can
also describe multi droplet geometries. The $\Lambda _{z},\Lambda
_{y},\Lambda _{x}$ records the information of dual giants, and in the
geometrical dual sense also the black droplets. $\Lambda _{z}^{c},\Lambda
_{y}^{c},\Lambda _{x}^{c}~$records the information of giants, and in the
geometrical dual sense also the white droplets. The white droplets are the
complementary region of black droplets, in the droplet space. They can be
viewed as the complement of the black droplets. Hence we can describe
various droplets in the higher dimensional droplet spaces, e.g. \cite%
{Chen:2007du}, see also e.g. \cite{Chong:2004ce}. One can also project the
higher dimensional droplets onto lower dimensional planes. There are also
matrix model descriptions \cite{Berenstein:2005aa}, see also e.g. \cite%
{Berenstein:2006yy}-\cite{Berenstein:2014isa}. It is also possible to have a
matrix model and collective field model combining $\Lambda _{z,y,x}$ with $%
\Lambda _{z,y,z}^{c}$, along the line of \cite{Berenstein:2022srd}.

\section{Reduced coherent states and fermionic counterparts}

\label{sec 6} \renewcommand{\theequation}{6.\arabic{equation}} %
\setcounter{equation}{0} \renewcommand{\thethm}{6.\arabic{thm}} %
\setcounter{thm}{0} \renewcommand{\theprop}{6.\arabic{prop}} %
\setcounter{prop}{0}

Here we consider reduced coherent states as well as their fermionic
counterparts. Starting from
\begin{equation}
\int dU\exp (\mathrm{tr}(\alpha U\Lambda U^{\dagger }Z)),
\end{equation}%
\cite{Holguin:2022zii} considered reduced BPS coherent states by considering
rank one projector $P_{1}$. Coherent states parameters are reduced from the $%
U(N)$ group average coherent state. In the convention here, we denote $%
\varphi _{1}=\varphi $. This is $N\times 1$ rectangular matrix. We have $%
\Lambda =\lambda P_{1}$ \cite{Holguin:2022zii} and
\begin{equation}
\varphi \varphi ^{\dagger }=UP_{1}U^{\dagger },~~~~~~U\Lambda U^{\dagger
}=\lambda \varphi \varphi ^{\dagger }.
\end{equation}%
This effectively reduces one-matrix to one-number. The reduced coherent
state is then
\begin{equation}
\mathcal{N}\int_{\mathbb{CP}^{N-1}}d\varphi ^{\dagger }d\varphi ~e^{\alpha
\lambda \varphi ^{\dagger }Z\varphi }.
\end{equation}%
Here the normalization factor is $\mathcal{N=}\frac{1}{\mathtt{Vol}(\mathbb{%
CP}^{N-1})}$. The $\alpha $ may be absorbed into the the redefinition of $%
\lambda $ or $\Lambda $.

We add fermion. Here $\chi $ is $N\times 1$ fermionic vector. Imagine we add
fermion counterpart $\chi $, so $\Psi =(\varphi ,\chi )$ locally parametrize
$C^{1|1}\otimes C^{N}$. Fermions are anti-commuting variables, see for
example \cite{Berezin,Berezin:87}, with e.g. $\chi _{i}^{\dagger \alpha
}\chi _{j\beta }=-\chi _{j\beta }\chi _{i}^{\dagger \alpha }$. For $p=1$,
the extended state is
\begin{equation}
\mathcal{N}\int d\varphi ^{\dagger }d\varphi d\chi ^{\dagger }d\chi e^{%
\mathrm{tr}(\chi \chi ^{\dagger })}\exp (\alpha \lambda \varphi ^{\dagger
}Z\varphi +\beta \lambda ^{c}\chi ^{\dagger }Z\chi ).
\end{equation}%
$\mathbb{CP}^{N-1|N}$ is co-dim one hypersurface in $C^{N|N}$. These kinds
of integral can be viewed as integration on curved supermanifold. As a
side-remark, the methods of integration over curved supermanifolds, e.g.
\cite{DeWitt,Witten:2012bh,Bettadapura:2020apr}, have also occurred in and
are very important for obtaining superstring amplitudes \cite{Witten:2012bh}.

Here $Z$ is a field coupling both the boson and fermion. $Z$ is finnally
dynamic with its own actions. The above is a operator with auxiliary field.
This describe states which are composed of a giant parametrized by $\lambda
^{c}$ and a dual giant parametrized by $\lambda $.

As pointed out in \cite{Holguin:2022zii}, one can consider rank $p$
projector $P_{p}$. Here $p=p_{1},~\varphi $ are\ $N\times p_{1}$, and$~\chi $
are $N\times p_{2}$ fermionic vectors. Denote $\Lambda _{p}$ as $p\times p$
block in $\Lambda $. Here $P_{p}$ is rank-$p$ projector,
\begin{equation}
\varphi \varphi ^{\dagger }=UP_{p}U^{\dagger },~~~~~\ U\Lambda U^{\dagger
}=\varphi \Lambda _{p}\varphi ^{\dagger },\ \
\end{equation}%
hence $\mathrm{tr}(\alpha U\Lambda U^{\dagger }Z)=\alpha \mathrm{tr}_{p_{1}}%
\mathrm{(}\Lambda _{p_{1}}\varphi ^{\dagger }Z\varphi )$. Then we also add
fermion. $\Lambda _{p_{2}}^{c}$ is a rank-$p_{2}$ complementary coherent
state parameter matrix. The superscript $c$ means or refers to
complementary. We have the extended state
\begin{equation}
\mathcal{N}\int d\varphi ^{\dagger }d\varphi d\chi ^{\dagger }d\chi e^{%
\mathrm{tr}(\chi \chi ^{\dagger })}\exp (\alpha \mathrm{tr}_{p_{1}}(\Lambda
_{p_{1}}\varphi ^{\dagger }Z\varphi )+\beta \mathrm{tr}_{p_{2}}(\Lambda
_{p_{2}}^{c}\chi ^{\dagger }Z\chi )).
\end{equation}%
They are reduced coherent state. The integral is reduced from the full
integral of the original group integral. The integration is on subspace of
the original group manifold. The correlation functions involving the reduced
coherent states describing the giant gravitons have been computed and agree
with gravity exactly \cite{Holguin:2022zii}.

We now discuss adding strings, and insertion of other operators, in the
reduced coherent states. $W$ is a word, or a string of letters. $\bar{\chi}%
W\chi $ adds open strings on giants, and $\bar{\varphi}W\varphi $ adds open
strings on dual giants. For example, $W_{l}$ is a string, e.g. $W_{l}=ZYX..$%
. Hence, with strings added, we have the states%
\begin{eqnarray}
&&\mathcal{N}\int d\varphi ^{\dagger }d\varphi d\chi ^{\dagger }d\chi e^{%
\mathrm{tr}(\chi \chi ^{\dagger })}\exp (\alpha \lambda \varphi ^{\dagger
}Z\varphi +\beta \lambda ^{c}\chi ^{\dagger }Z\chi )  \notag \\
&&(\varphi ^{\dagger }W_{1}\varphi )(\varphi ^{\dagger }W_{2}\varphi
)...(\varphi ^{\dagger }W_{k}\varphi )  \notag \\
&&(\chi ^{\dagger }W_{1}^{\prime }\chi )(\chi ^{\dagger }W_{2}^{\prime }\chi
)...(\chi ^{\dagger }W_{l}^{\prime }\chi ).
\end{eqnarray}%
More generally, for $\int d\varphi ^{\dagger }d\varphi d\chi ^{\dagger
}d\chi e^{\mathrm{tr}(\chi \chi ^{\dagger })}=\int \prod_{i,j}d\varphi
_{i}^{\dagger }d\varphi _{i}d\chi _{j}^{\dagger }d\chi _{j}e^{\mathrm{tr}%
(\chi \chi ^{\dagger })}$, where $1\leqslant i,j\leqslant p_{1},p_{2}$, we
have the string-added states
\begin{eqnarray}
&&\mathcal{N}\int d\varphi ^{\dagger }d\varphi d\chi ^{\dagger }d\chi e^{%
\mathrm{tr}(\chi \chi ^{\dagger })}\exp (\alpha \mathrm{tr}_{p_{1}}\mathrm{(}%
\Lambda _{p_{1}}\varphi ^{\dagger }Z\varphi )+\beta \mathrm{tr}%
_{p_{2}}(\Lambda _{p_{2}}^{c}\chi ^{\dagger }Z\chi ))  \notag \\
&&(\varphi _{i_{1}}^{\dagger }W_{1}\varphi _{j_{1}})(\varphi
_{i_{2}}^{\dagger }W_{2}\varphi _{j_{2}})...(\varphi _{i_{k}}^{\dagger
}W_{k}\varphi _{j_{k}})  \notag \\
&&(\chi _{i_{1}}^{\dagger }W_{1}^{\prime }\chi _{j_{1}})(\chi
_{i_{2}}^{\dagger }W_{2}^{\prime }\chi _{j_{2}})...(\chi _{i_{l}}^{\dagger
}W_{l}^{\prime }\chi _{j_{l}}).
\end{eqnarray}%
This way of adding or inserting strings is essentially equivalent to that of
\cite{Berenstein:2022srd,Lin:2022wdr}.

We discuss reduced coherent states in quarter BPS and eighth BPS case. $%
Z,Y,X $ are coupled in the same way with $U$. We have the extended states
\begin{eqnarray}
&&\int d\varphi ^{\dagger }d\varphi d\chi ^{\dagger }d\chi e^{\mathrm{tr}%
(\chi \chi ^{\dagger })}\exp (\alpha \mathrm{tr}_{p_{1}}(\Lambda
_{zp_{1}}\varphi ^{\dagger }Z\varphi +\Lambda _{yp_{1}}\varphi ^{\dagger
}Y\varphi +\Lambda _{xp_{1}}\varphi ^{\dagger }X\varphi )  \notag \\
&&+\beta \mathrm{tr}_{p_{2}}(\Lambda _{zp_{2}}^{c~}\chi ^{\dagger }Z\chi
+\Lambda _{yp_{2}}^{c}\chi ^{\dagger }Y\chi +\Lambda _{xp_{2}}^{c}\chi
^{\dagger }X\chi ).
\end{eqnarray}%
The $\Lambda _{p_{2}}^{c}$ plays the role of submatrix of the complementary
coherent state parameter $\Lambda _{z}^{c}$ of Sec. 2.

The operators have rich sub-block structures. Note that the non-zero entries
of $\Lambda _{(z,y,x)p_{2}}^{c}$ can be in orthogonal subspaces. Consider
special cases. They have sub-diagonal-blocks in orthogonal subspaces, e.g.
schematically for the relevant part (the $\Lambda _{m}^{c~}s\ $below are
diagonal for simplicity),
\begin{equation}
\Lambda _{zp_{2}}^{c~}=\left[
\begin{array}{ccc}
\Lambda _{m_{21}}^{c~} &  &  \\
& 0 &  \\
&  & 0%
\end{array}%
\right] ,\Lambda _{yp_{2}}^{c~}=\left[
\begin{array}{ccc}
0 &  &  \\
& \Lambda _{m_{22}}^{c~} &  \\
&  & 0%
\end{array}%
\right] ,\Lambda _{xp_{2}}^{c~}=\left[
\begin{array}{ccc}
0 &  &  \\
& 0 &  \\
&  & \Lambda _{m_{23}}^{c~}%
\end{array}%
\right] .
\end{equation}%
The states reduce to
\begin{equation}
\prod_{i=1}^{m_{21}}\det \left[ \lambda _{zi}^{c}Z-\beta ^{-1}\right]
\prod_{i=1}^{m_{22}}\det \left[ \lambda _{yi}^{c}Y-\beta ^{-1}\right]
\prod_{i=1}^{m_{23}}\det \left[ \lambda _{xi}^{c}X-\beta ^{-1}\right] .
\end{equation}%
There are other related states. If the sub-blocks are in parallel subspaces,
e.g. schematically for the relevant part,
\begin{equation}
\Lambda _{zp_{2}}^{c~}=\left[
\begin{array}{ccc}
0 &  &  \\
& \Lambda _{m_{21}}^{c~} &  \\
&  & 0%
\end{array}%
\right] ,\Lambda _{yp_{2}}^{c~}=\left[
\begin{array}{ccc}
0 &  &  \\
& \Lambda _{m_{22}}^{c~} &  \\
&  & 0%
\end{array}%
\right] ,\Lambda _{xp_{2}}^{c~}=\left[
\begin{array}{ccc}
0 &  &  \\
& 0 &  \\
&  & \Lambda _{m_{23}}^{c~}%
\end{array}%
\right] ,
\end{equation}%
the states reduce to
\begin{equation}
\prod_{i=1}^{m_{21}}\det \left[ \lambda _{zi}^{c}Z+\lambda _{yi}^{c}Y-\beta
^{-1}\right] \prod_{i=1}^{m_{23}}\det \left[ \lambda _{xi}^{c}X-\beta ^{-1}%
\right] .
\end{equation}%
The two types of operators are related by a change of the sub-structure of
the coherent state parameter matrix. This illustrates the rich patterns of
sub-block structures. This leads also to a huge degeneracy of microstates.

\section{Discussion}

\label{sec_discussion}

We generalized a construction of coherent state operators describing various
giant graviton branes. We enlarged the coherent state parameters, by
including complementary coherent state parameters $\Lambda ^{c}$, to
describe a system of at most $N$ dual giants and at most $T$ giants. The
methods in this paper is complementary to related works in the literature.

We treat giants and dual giants on the equal footing, and make the roles of $%
N$ and of $T$, symmetric. This is evident in the construction of fermionic
extended coherent states where we have both $N\times N$ and $T\times T$
coherent state amplitudes describing giants and dual giants. One of the
advantages of using complementary $T\times T$ coherent state parameters is
that they have rich sub-block structures that record different type of giant
graviton states, with different numbers of giants and different momenta. We
packaged the coherent state parameters into supermatrix. We use supermatrix
to construct special coherent states, which encode information of both
giants and dual giants.

We find and construct states that interpolate both the group integral
coherent states and multi determinantal states, as well as schur states. We
also built two auxiliary models as intermediate steps of constructing
operators. In some sense, we have made unified treatments of these
operators. The construction and generalization of these operators \cite%
{Berenstein:2022srd} are useful for the moduli-space of a system of giant
gravitons \cite{Biswas:2006tj,Mandal:2006tk,Pasukonis:2012zj}.

The near-BPS states can be considered as excitations on background BPS
states. Hence the constructions of BPS operators are useful for analyzing
near-BPS states. It is convenient to add strings. The coherent state is BPS,
hence we can extract the near BPS spectrum subtracted from the BPS coherent
state background. The string-added states capture near-BPS states. The
operators are closely relevant for the calculations of excitations of giant
graviton branes, as well as excitations of bubbling geometries. Near BPS
spectrum of giants can also be captured by restricted Schurs, e.g. \cite%
{deMelloKoch:2011wah,Carlson:2011hy,deMelloKoch:2012ck,Lin:2014yaa}, and by
spin matrix theory, e.g. \cite{Harmark:2016cjq,Baiguera:2021hky}. This is
also in accord with the ideas and objectives in spin matrix theory. It is
also convenient for calculating the interactions of giant graviton branes
with strings.

When calculating the physical observables, the auxiliary variables should be
integrated out finally. They can be kept in intermediate steps, and due to
their presence, the intermediate-step calculations can be simpler. The
reason is that the intermediate step involves coherent states, which have
more convenient algebraic manipulations. For example, we can first calculate
the correlation function of un-integrated operators, then we make multi
auxiliary field integrations, to obtain the correlation function of
integrated operators. The coherent state representation, auxiliary integrals
and susy integrals, facilitate the computations efficiently. The coherent
state representation facilitates the calculation of ladder operators and
have the advantage of simplifying the action of the involved dilatation
operators.

Very recently, very interesting and closely related analysis have also
appeared in \cite{Liu:2022olj} and \cite{Carlson:2022dot}, among other
things. Adding strings in a different class of operators, the restricted
Schur operators, instead of in the coherent state operators, have also
appeared very recently \cite{Carlson:2022dot}. Among other things, this work
includes the restricted-Schur counterpart of the $T=2$ case, and have been
recently worked out. The underlying operators in both cases are closely
related. The coherent state operators and restricted Schur operators use
different variables and parameters, and the mappings of the parameters
between the two different classes of the operators are nontrivial. Adding
strings are very convenient for both classes of operators.

The centrally-extended $SU(2|2)$ symmetry plays important roles in
simplifying the properties of spectrum and S-matrix of exited states. The
string-added states capture near-BPS states. Since the coherent state is
itself BPS, we can extract the near BPS spectrum subtracted from the BPS
coherent state background. The excitations of the states are expected to
include giant magnons, spinning strings and other strings, as well as in
close connection to $SU(2|2)$ symmetry and $SL(2)$ sectors, e.g. \cite%
{Berenstein:2020jen,Kim:2018gwx,Tseytlin}.

The giant configurations are also related to near-extremal black holes. Near
BPS states can also describe near BPS black holes, and are related to
various giant configurations and intersecting giants. The string
configurations \cite{Berenstein:2022cju} on giants are relevant. A better
understanding of the quarter and eighth BPS sectors as well as near-BPS
sectors will give implications for physics of extremal and near-extremal
black holes e.g. \cite{Balasubramanian:2007bs,Fareghbal:2008ar}.

Three-point correlation functions of giant gravitons have been computed from
gravity and from gauge theory sides, and they match with each, e.g. \cite%
{Bissi:2011dc}-\cite{Kristjansen:2015gpa},\cite{Yang:2021kot},\cite%
{Holguin:2022zii}. The operators used in the computations are Schur
operators and their linear superpositions. The gravity computation involves
brane DBI action describing the interactions of branes with lighter probes.
It has been very useful computing three-point and four-point functions
involving these heavy large operators. The correlation functions involving
the above-described coherent state operators and lighter operators have also
been computed and agree precisely with gravity \cite{Holguin:2022zii}.

Coherent states, as well as multi-determinantal states and multi-schur
states, are closely related, and they are also dual to gravity geometries,
e.g. \cite{Berenstein:2017abm},\cite{Lin:2017dnz},\cite{Belin:2020zjb}-\cite%
{Berenstein:2017rrx}. Transformation between bases are useful for
computations, including those aspects related to amplitudes of gravity side.
These heavy operators involve emergent geometries in the dual quantum
gravity system, e.g. \cite{Grant:2005qc}-\cite{Balasubramanian:2007zt} and
related discussions. These excited states live in the same Hilbert space of
the gravity side. Since they live in the same Hilbert space, we can
dynamically relate them using the Hamiltonian in the same Hilbert space.

This is also suggestive that there can be a large $N$ eigenvalue density
approach. The dynamics and evolution of eigenvalues $\lambda (t)$ can be
described using collective field methods, e.g. \cite{Berenstein:2013md}. The
evolution of eigenvalues in some regimes should be related to observations
in \cite{deMelloKoch:2020jmf}. There are also various integrability breaking
deformations and perturbations that can be added into these systems. There
are rich sub-block structures describing various brane configurations. These
scenarios are also closely related to blackhole microstate properties and
fuzzball microstructures, e.g. \cite{Mathur:2005ai}-\cite{Mayerson:2020tpn}.

There are also interesting and closely related phenomena in Wilson line
operators. The involved path integral can be computed by saddle point method
and localization. The localization techniques are also very useful and have
occurred for Wilson operators. Furthermore, emergent geometries are also
dual to large Wilson loop operators, e.g. \cite{Yamaguchi:2006te}-\cite%
{Gomis:2008qa} as examples. It would be nice to have an unified
understanding together with these situations.

Different operator bases can be transformed to each other, e.g. \cite%
{Lin:2017vfn}. There are also coherent states constructed using Brauer
algebras \cite{Lin:2017vfn}, in relation to \cite{Kimura:2007wy}, with
another operator basis. On the other hand, there are other representation
bases of operators, including Brauer operators, flavor symmetry bases, and
multi restricted Schur operators \cite{Kimura:2007wy}-\cite{Brown:2008ij}.
It would be interesting to understand more precise relations between them
and the coherent state operators, and between these classes of operators.

\section*{Acknowledgments}

We would like to thank A. Belin, B. Czech, J. Hou, R. de Mello Koch, S.
Komatsu, L.Y. Hung, Y. Jiang, Z.J. Li, S. Ramgoolam, J. Simon, M. Sperling,
R. Suzuki, Q. Wen, J.B. Wu, P. Yang, R.D. Zhu for related discussions. The
work was supported in part by National Key R\&D Program of China grant No.
2020YFA0713000, by Overseas high-level talents program, and by Fundamental
Research Funds for the Central Universities of China.

\appendix

\section{Inner products of purely fermionic model}

\label{sec A} \renewcommand{\theequation}{A.\arabic{equation}} %
\setcounter{equation}{0} \renewcommand{\thethm}{A.\arabic{thm}} %
\setcounter{thm}{0} \renewcommand{\theprop}{A.\arabic{prop}} %
\setcounter{prop}{0}

We look at more details on
\begin{eqnarray}
O &=&\int d\chi ^{\dagger }d\chi e^{\mathrm{tr}(\chi \chi ^{\dagger })}\exp
(-\mathrm{tr}(\beta \chi \Lambda _{z}^{c}\chi ^{\dagger }Z+\beta \chi
\Lambda _{y}^{c}\chi ^{\dagger }Y+\beta \chi \Lambda _{x}^{c}\chi ^{\dagger
}X)) \\
&=&\int d\chi ^{\dagger }d\chi e^{\mathrm{tr}(\chi \chi ^{\dagger })}O_{0}.
\end{eqnarray}%
We have%
\begin{equation}
G=\int d\chi ^{\prime \dagger }d\chi ^{\prime }e^{\mathrm{tr}(\chi ^{\prime
}\chi ^{\prime \dagger })}\langle O_{0}|O_{0}\rangle .
\end{equation}%
We illustrate this idea or method by the simpler case
\begin{eqnarray}
O &=&\int d\chi ^{\dagger }d\chi e^{\mathrm{tr}(\chi \chi ^{\dagger })}\exp
(-\mathrm{tr}(\beta \chi \Lambda _{z}^{c}\chi ^{\dagger }Z)) \\
&=&\int d\chi ^{\dagger }d\chi e^{\mathrm{tr}(\chi \chi ^{\dagger })}O_{0}
\end{eqnarray}%
and we have%
\begin{equation}
\langle O|O\rangle =\int d\chi ^{\dagger }d\chi e^{\mathrm{tr}(\chi \chi
^{\dagger })}G~.
\end{equation}%
The advantage is that the un-integrated operators are coherent state
operators, which have nice behavior under the annihilation operators.%
\begin{eqnarray}
\langle O_{0}|O_{0}\rangle  &\sim &\mathcal{N}\exp (\mathrm{tr}_{N}(\beta
\chi \Lambda _{z}^{c}\chi ^{\dagger }\beta \chi ^{\prime }\Lambda
_{z}^{c\dagger }\chi ^{\prime \dagger }))  \notag \\
&\sim &\mathcal{N}\exp (-\mathrm{tr}_{T}(\beta ^{2}\chi ^{\prime \dagger
}\chi \Lambda _{z}^{c}\chi ^{\dagger }\chi ^{\prime }\Lambda _{z}^{c\dagger
}))  \notag \\
&\sim &\mathcal{N}\exp (-\sum_{i}(\beta ^{2}\lambda _{zi}^{c\dagger }\chi
_{i}^{\prime \dagger }\left( \chi \Lambda _{z}^{c}\chi ^{\dagger }\right)
\chi _{i}^{\prime })).
\end{eqnarray}%
The below is for when $T=N$.%
\begin{eqnarray}
G &\sim &\int d\chi ^{\prime \dagger }d\chi ^{\prime }e^{\mathrm{tr}(\chi
^{\prime }\chi ^{\prime \dagger })}\langle O_{0}|O_{0}\rangle   \notag \\
&\sim &\mathcal{N}\int d\chi ^{\prime \dagger }d\chi ^{\prime }e^{-\mathrm{tr%
}(\chi ^{\prime \dagger }\chi ^{\prime })}\exp (-\sum_{i}(\chi _{i}^{\prime
\dagger }\left( \beta ^{2}\lambda _{zi}^{c\ast }\chi \Lambda _{z}^{c}\chi
^{\dagger }\right) \chi _{i}^{\prime }))  \notag \\
&\approx &\mathcal{N}\prod_{i}\det {}\left( 1+\beta ^{2}\lambda _{zi}^{c\ast
}\chi \Lambda _{z}^{c}\chi ^{\dagger }\right)   \notag \\
&\approx &\mathcal{N}\det {}\left( 1+\beta ^{2}\Lambda _{z}^{c\dagger }\chi
\Lambda _{z}^{c}\chi ^{\dagger }\right) .
\end{eqnarray}%
Similarly, we can calculate for two states with different amplitudes. $%
\mathcal{N}$ is only a constant numerical factor, which is a overal scale
factor related to $\beta $, $N$, etc, and can be restored by dimensional
counting.

\section{Generalization to giant-anti-giant related system}

\label{sec B} \renewcommand{\theequation}{B.\arabic{equation}} %
\setcounter{equation}{0} \renewcommand{\thethm}{B.\arabic{thm}} %
\setcounter{thm}{0} \renewcommand{\theprop}{B.\arabic{prop}} %
\setcounter{prop}{0}

We include also $\bar{Z},{\bar{Y},\bar{X}}$ and the states are more general:%
\begin{equation}
\Lambda _{z}^{S}=\left(
\begin{array}{cc}
\alpha \Lambda _{z} & 0 \\
0 & -\beta \Lambda _{z}^{c}%
\end{array}%
\right) ,\Lambda _{y}^{S}=\left(
\begin{array}{cc}
\alpha \Lambda _{y} & 0 \\
0 & -\beta \Lambda _{y}^{c}%
\end{array}%
\right) ,\Lambda _{x}^{S}=\left(
\begin{array}{cc}
\alpha \Lambda _{x} & 0 \\
0 & -\beta \Lambda _{x}^{c}%
\end{array}%
\right) .
\end{equation}%
\begin{eqnarray}
Z_{S} &=&\left(
\begin{array}{cc}
Z & 0 \\
0 & \bar{Z}%
\end{array}%
\right) ,Y_{S}=\left(
\begin{array}{cc}
{Y} & 0 \\
0 & {\bar{Y}}%
\end{array}%
\right) ,X_{S}=\left(
\begin{array}{cc}
{X} & 0 \\
0 & {\bar{X}}%
\end{array}%
\right) . \\
I &=&\int [dU_{S}]\exp (\mathrm{Str}(\Lambda _{z}^{S}U_{S}^{\dagger
}Z_{S}U_{S}+\Lambda _{y}^{S}U_{S}^{\dagger }Y_{S}U_{S}+\Lambda
_{x}^{S}U_{S}^{\dagger }X_{S}U_{S})).
\end{eqnarray}%
These are the operators $O_{1}[\alpha \Lambda _{z,y,x},-\beta \Lambda
_{z,y,x}^{c};Z,\bar{Z},{Y,\bar{Y},X,\bar{X}}]$. This can be viewed as a
generalization of (\ref{op 13}),(\ref{op 15}) by including $\bar{Z},{\bar{Y},%
\bar{X}}$. For simplicity we write the case without $\Lambda _{x}^{S}$,
\begin{eqnarray}
&&O_{1}[\alpha \Lambda _{z,y},-\beta \Lambda _{z,y}^{c};Z,\bar{Z},{Y,\bar{Y}}%
]  \notag \\
&=&\int dUdU_{c}d\chi ^{\dagger }d\chi e^{\mathrm{tr}(\chi \chi ^{\dagger
})}\exp (\mathrm{tr}_{N}(\alpha U\Lambda _{z}U^{\dagger }Z+\alpha \chi
^{\dagger }\Lambda _{z}\chi \bar{Z}+\alpha U\Lambda _{y}U^{\dagger }Y+\alpha
\chi ^{\dagger }\Lambda _{y}\chi {\bar{Y}}  \notag \\
&&+\beta U_{c}\Lambda _{z}^{c}U_{c}^{\dagger }\bar{Z}-\beta \chi \Lambda
_{z}^{c}\chi ^{\dagger }Z+\beta U_{c}\Lambda _{y}^{c}U_{c}^{\dagger }{\bar{Y}%
}-\beta \chi \Lambda _{y}^{c}\chi ^{\dagger }Y)).
\end{eqnarray}%
Here we refer to $\det \bar{Z},\det {\bar{Y},\det \bar{X}}$ as anti-giants.
These are related to giant-anti-giant systems. This state also contains $%
SO(4)$ singlets and $SO(6)$ singlets, inside it. It might be nice to
understand their relation to Brauer operators \cite{Kimura:2007wy}.

\section{More detailed properties of $O_{2}$}

\label{sec C} \renewcommand{\theequation}{C.\arabic{equation}} %
\setcounter{equation}{0} \renewcommand{\thethm}{C.\arabic{thm}} %
\setcounter{thm}{0} \renewcommand{\theprop}{C.\arabic{prop}} %
\setcounter{prop}{0}

Now consider $O_{2}[\alpha \Lambda _{z},-\beta \Lambda _{z}^{c};Z,Z]$ and it
is half BPS.
\begin{eqnarray}
&&O_{2}[\alpha \Lambda _{z},-\beta \Lambda _{z}^{c};Z,Z]  \notag \\
&=&\int dU_{c}dUd\chi ^{\dagger }d\chi e^{\mathrm{tr}(\chi \chi ^{\dagger
})}\exp (\mathrm{tr}(\alpha U\Lambda _{z}U^{\dagger }Z-\beta \chi \Lambda
_{z}^{c}\chi ^{\dagger }Z+\beta U_{c}\Lambda _{z}^{c}U_{c}^{\dagger
}Z+\alpha \chi ^{\dagger }\Lambda _{z}\chi Z)).  \notag \\
&&
\end{eqnarray}

For the moment, we make a simplified case that we assume now that $\Lambda
=\Lambda _{z}$ is diagonal. The fermions are anti-commuting, hence $\mathrm{%
tr}(\alpha \chi ^{\dagger }\Lambda \chi Z-\beta \chi \Lambda ^{c}\chi
^{\dagger }Z)$ = $-\mathrm{tr}(\chi (\alpha \Lambda +\beta \Lambda ^{c})\chi
^{\dagger }Z)$. Hence%
\begin{eqnarray}
&&O_{2}[\alpha \Lambda ,-\beta \Lambda ^{c};Z,Z]  \notag \\
&\sim &\int dU_{c}dUd\chi ^{\dagger }d\chi e^{\mathrm{tr}(\chi \chi
^{\dagger })}\exp (\mathrm{tr}(\alpha U\Lambda U^{\dagger }Z+\beta
U_{c}\Lambda ^{c}U_{c}^{\dagger }Z+(\alpha \Lambda +\beta \Lambda ^{c})\chi
^{\dagger }Z\chi )).  \notag \\
&&
\end{eqnarray}

We now also make an approximation that we assume the special case that $%
\Lambda ^{c}$ is diagonal. These are not the most general cases.%
\begin{eqnarray}
&&O_{2}[\alpha \Lambda ,-\beta \Lambda ^{c};Z,Z]  \notag \\
&\sim &\int dU_{c}dU\exp (\mathrm{tr}(\alpha U\Lambda U^{\dagger }Z+\beta
U_{c}\Lambda ^{c}U_{c}^{\dagger }Z))  \notag \\
&&{\prod_{i}}\det (\left( \alpha \lambda _{i}+\beta \lambda _{i}^{c}\right)
Z-1)~.\ \
\end{eqnarray}%
The norm of the operator is%
\begin{eqnarray}
&&||O_{2}[\alpha \Lambda ,-\beta \Lambda ^{c};Z,Z]||^{2}  \notag \\
&\approx &\mathcal{N~}{\prod_{i,j}}(\alpha \lambda _{i}+\beta \lambda
_{j}^{c})(\alpha \bar{\lambda}_{i}+\beta \bar{\lambda}_{j}^{c})\exp
(\sum\limits_{i,j}(\left( \alpha \lambda _{i}+\beta \lambda _{j}^{c}\right)
(\alpha \bar{\lambda}_{i}+\beta \bar{\lambda}_{j}^{c}))^{-1})  \notag \\
&&\frac{\det \left( \exp (\alpha ^{2}\bar{\lambda}_{i}\lambda _{j})\right) }{%
\Delta (\alpha \bar{\lambda})\Delta (\alpha \lambda )}\frac{\det \left( \exp
(\beta ^{2}\bar{\lambda}_{i}^{c}\lambda _{j}^{c})\right) }{\Delta (\beta
\bar{\lambda}^{c})\Delta (\beta \lambda ^{c})}.
\end{eqnarray}%
$\mathcal{N}$ is eigenvalue-independent combinatoric factor, and can be
restored by dimensional analyses of $\alpha ,\beta $, and $\Delta (\alpha
\lambda )$ is the Vandemonde determinant.

In the $\beta \Lambda ^{c}\rightarrow 0~$limit, it reduces to
\begin{eqnarray}
&&||O_{2}[\alpha \Lambda ,0;Z,Z]||^{2}\approx \mathcal{N~}{\prod_{i}}(\alpha
^{2}\bar{\lambda}_{i}\lambda _{i})^{N}\exp (N\sum\limits_{i}(\alpha ^{-2}(%
\bar{\lambda}_{i}\lambda _{i})^{-1}))\frac{\det \left( \exp (\alpha ^{2}\bar{%
\lambda}_{i}\lambda _{j})\right) }{\Delta (\alpha \bar{\lambda})\Delta
(\alpha \lambda )}.  \notag \\
&&
\end{eqnarray}%
The $\alpha ^{2}\bar{\lambda}_{i}\lambda _{i}$ factors are from rescaling of
$\det (\alpha \lambda _{i}Z-1)\rightarrow \det (Z-(\alpha \lambda
_{i})^{-1}) $. This agrees with previous results e.g. \cite%
{Berenstein:2022srd,Holguin:2022zii,Chen:2019gsb,Corley:2001zk}. This is
also suggestive that there can be a large $N$ eigenvalue density approach.

Our integrand is closely related to HCIZ and in particular super HCIZ, e.g.
\cite{Harish}-\cite{Alfaro:1994ca}. The techniques of these methods might be
useful in some special cases.

\section{More general superunitary related integral}

\label{sec D} \renewcommand{\theequation}{D.\arabic{equation}} %
\setcounter{equation}{0} \renewcommand{\thethm}{C.\arabic{thm}} %
\setcounter{thm}{0} \renewcommand{\theprop}{D.\arabic{prop}} %
\setcounter{prop}{0}

We discuss some other super integrals which are related to the HCIZ type,
and are related to the constructions of operators.

Consider
\begin{equation}
I=\int [dU_{S}]\text{ \textrm{sdet}}^{-1}\left( I_{S}-\Lambda
_{S}U_{S}^{\dagger }Z_{S}U_{S}\right) .
\end{equation}%
Here $Z_{S}=\left(
\begin{array}{cc}
Z & 0 \\
0 & 0%
\end{array}%
\right) ,\Lambda _{S}=\left(
\begin{array}{cc}
\alpha \Lambda & 0 \\
0 & -\beta \Lambda ^{c}%
\end{array}%
\right) $. $\Lambda ^{c}$ is $T\times T$ and $\Lambda $ is $N\times N$. In
component form,
\begin{equation}
O\sim \int dUd\chi ^{\dagger }d\chi e^{\mathrm{tr}(\chi \chi ^{\dagger })}%
\frac{\det_{N}(I-\beta \chi \Lambda ^{c}\chi ^{\dagger }Z)}{%
\det_{N}(I-\alpha U\Lambda U^{\dagger }Z)}.
\end{equation}%
Note e.g. $\det_{T}(I+\beta \Lambda ^{c}\chi ^{\dagger }Z\chi
)=\det_{N}(I-\beta \chi \Lambda ^{c}\chi ^{\dagger }Z)$, because fermions
are anti-commuting variables.

We generalize it to%
\begin{equation}
O\sim \int [dU_{S}]~\mathrm{sdet}^{-1}\left( I_{S}-\Lambda
_{z}^{S}U_{S}^{\dagger }Z_{S}U-\Lambda _{y}^{S}U_{S}^{\dagger
}Y_{S}U_{S}-\Lambda _{x}^{S}U_{S}^{\dagger }X_{S}U_{S}\right) .
\end{equation}%
In component form,%
\begin{equation}
O\sim \int dUd\chi ^{\dagger }d\chi e^{\mathrm{tr}(\chi \chi ^{\dagger })}%
\frac{\det_{N}(I-\beta \chi \Lambda _{z}^{c}\chi ^{\dagger }Z-\beta \chi
\Lambda _{y}^{c}\chi ^{\dagger }Y-\beta \chi \Lambda _{x}^{c}\chi ^{\dagger
}X)}{\det_{N}(I-\alpha U\Lambda _{z}U^{\dagger }Z-\alpha U\Lambda
_{y}U^{\dagger }Y-\alpha U\Lambda _{x}U^{\dagger }X)}.  \label{operators 17}
\end{equation}%
This is for general $T,N$. We can first compute the norm of the integrand,
and then do double integration of auxiliary variables. This idea is parallel
to double integral of norm, which was referred to as convolution \cite%
{Berenstein:2022srd}.

For $T=N$, we can also use a more complicated version $Z_{S}=\left(
\begin{array}{cc}
Z & 0 \\
0 & Z%
\end{array}%
\right) $ and similarly for $Y_{S},X_{S}$, then there is extra factor $%
\det_{N}(I+\beta U\Lambda _{z}^{c}U^{\dagger }Z+\beta U\Lambda
_{z}^{c}U^{\dagger }Y+\beta U\Lambda _{z}^{c}U^{\dagger }X)~$in the
numerator, similar to the denominator in (\ref{operators 17}).

The expansion in the universal limit is%
\begin{eqnarray}
&&\int dUd\chi ^{\dagger }d\chi e^{\mathrm{tr}(\chi \chi ^{\dagger })}\exp (%
\mathrm{tr}_{N}(\alpha U\Lambda _{z}U^{\dagger }Z-\beta \chi \Lambda
_{z}^{c}\chi ^{\dagger }Z+\alpha U\Lambda _{y}U^{\dagger }Y-\beta \chi
\Lambda _{y}^{c}\chi ^{\dagger }Y  \notag \\
&&+\alpha U\Lambda _{x}U^{\dagger }X-\beta \chi \Lambda _{x}^{c}\chi
^{\dagger }X)).~~~~~~
\end{eqnarray}

Now we consider the parity-transposition operations, which switch the BB
with FF, and are related to giant-dual-giant transformations. The
parity-transposition transformed variables are $Z_{S}^{\pi }=\left(
\begin{array}{cc}
0 & 0 \\
0 & Z%
\end{array}%
\right) $, $\Lambda _{z}^{S\pi }=\left(
\begin{array}{cc}
-\beta \Lambda _{z}^{c} & 0 \\
0 & \alpha \Lambda _{z}%
\end{array}%
\right) $, and also for $\Lambda _{y}^{S\pi }$, $\Lambda _{x}^{S\pi }$, $%
Y_{S}^{\pi }$, $X_{S}^{\pi }$. The parity-transposition transformed
operators $O_{\pi }$ is%
\begin{eqnarray}
&\sim &\int [dU_{S}]\text{ \textrm{sdet}}^{-1}\left( I_{S}-\Lambda
_{z}^{S\pi }U_{S}^{\dagger }Z_{S}^{\pi }U_{S}-\Lambda _{y}^{S\pi
}U_{S}^{\dagger }Y_{S}^{\pi }U_{S}-\Lambda _{x}^{S\pi }U_{S}^{\dagger
}X_{S}^{\pi }U_{S}\right)  \notag \\
&\sim &\int [dU_{S}]\text{ \textrm{sdet}}\left( I_{S}-\Lambda
_{z}^{S}U_{S}^{\dagger }Z_{S}U_{S}-\Lambda _{y}^{S}U_{S}^{\dagger
}Y_{S}U_{S}-\Lambda _{x}^{S}U_{S}^{\dagger }X_{S}U_{S}\right) ,
\end{eqnarray}%
because the superdeterminant flips itself under the parity-transposition.
Comparing the two, they look inverse to each other. In component form,%
\begin{equation}
O_{\pi }=\int dUd\chi d\chi ^{\dagger }e^{\mathrm{tr}(\chi ^{\dagger }\chi )}%
\frac{\det_{N}(I-\alpha U\Lambda _{z}U^{\dagger }Z-\alpha U\Lambda
_{y}U^{\dagger }Y-\alpha U\Lambda _{x}U^{\dagger }X)}{\det_{N}(I-\beta \chi
\Lambda _{z}^{c}\chi ^{\dagger }Z-\beta \chi \Lambda _{y}^{c}\chi ^{\dagger
}Y-\beta \chi \Lambda _{x}^{c}\chi ^{\dagger }X)}.
\end{equation}%
In other words, determinant in the denominator can be flipped to the
determinant in the numerator, by the parity-transposition. This hence can
change a dual-giant operator $O$ to a giant operator $O_{\pi }$. These can
already illustrate the giant-dual-giant duality-transformation.

\section{Generating correlators of integrated operators from correlators of
un-integrated operators}

\label{sec E} \renewcommand{\theequation}{E.\arabic{equation}} %
\setcounter{equation}{0} \renewcommand{\thethm}{E.\arabic{thm}} %
\setcounter{thm}{0} \renewcommand{\theprop}{E.\arabic{prop}} %
\setcounter{prop}{0}

This idea can also work for calculating three-point and four-point
functions. The norm is a special case of two-point functions. We first
calculate the $n$-point correlation function of un-integrated operators,
then we do multi auxiliary field integrations. For example, we first do norm
of unintegrated operator, then do double integration of the auxiliary
variables, to get the norm for integrated operators. The idea is essentially
similar to the convolution idea or double integral in \cite%
{Berenstein:2022srd}; There, the HCIZ was encountered in a later stage of
integrating out the auxiliary variables. One can calculate more inner
products and correlation functions, using these representations e.g. (\ref%
{operators}) and (\ref{operators 17}).

For example,
\begin{equation}
O_{0}(\chi ,\chi ^{\dagger },Z,Y,X)=\det {}_{N}(I+\beta \chi \Lambda
_{z}^{c}\chi ^{\dagger }Z+\beta \chi \Lambda _{y}^{c}\chi ^{\dagger }Y+\beta
\chi \Lambda _{x}^{c}\chi ^{\dagger }X).
\end{equation}%
We refer to this as un-integrated operator. The integrated operator is
\begin{eqnarray}
O &=&\int d\chi ^{\dagger }d\chi e^{\mathrm{tr}(\chi \chi ^{\dagger })}O_{0}
\\
&=&\int d\chi ^{\dagger }d\chi e^{\mathrm{tr}(\chi \chi ^{\dagger })}\det
{}_{N}(I+\beta \chi \Lambda _{z}^{c}\chi ^{\dagger }Z+\beta \chi \Lambda
_{y}^{c}\chi ^{\dagger }Y+\beta \chi \Lambda _{x}^{c}\chi ^{\dagger }X).
\end{eqnarray}%
We refer to this as integrated operator.

We can also consider%
\begin{equation}
\det {}_{N}(I+\beta \chi \Lambda _{z}^{c}\chi ^{\dagger }Z+\beta \chi
\Lambda _{y}^{c}\chi ^{\dagger }Y+\beta \chi \Lambda _{x}^{c}\chi ^{\dagger
}X)^{-1}~,
\end{equation}%
as this can also appear as in Appendix D.

We have un-integrated operators $O_{0i}(\chi ,\chi ^{\dagger },Z,Y,X;W)$ and
integrated operators $O_{i}$. Here $W$ collectively denotes other fields.%
\begin{equation}
O_{i}=\int d\chi ^{\dagger }d\chi e^{\mathrm{tr}(\chi \chi ^{\dagger
})}O_{0i}(\chi ,\chi ^{\dagger },Z,Y,X;W).
\end{equation}%
Before the integration, it is easier to compute the correlation functions of
un-integrated operators, e.g. of the form$~\langle O_{04}^{\dagger
}O_{03}^{\dagger }O_{02}O_{01}\rangle $, $\langle O_{04}^{\dagger
}O_{02}^{\dagger }O_{03}O_{01}\rangle $ or $\langle O_{03}^{\dagger
}O_{02}^{\dagger }O_{04}O_{01}\rangle $. We have%
\begin{equation}
G(\chi ,\chi ^{\dagger })=\int d\chi ^{\prime \prime \prime \dagger }d\chi
^{\prime \prime \prime }e^{\mathrm{tr}(\chi ^{\prime \prime \prime }\chi
^{\prime \prime \prime \dagger })}\int d\chi ^{\prime \prime \dagger }d\chi
^{\prime \prime }e^{\mathrm{tr}(\chi ^{\prime \prime }\chi ^{\prime \prime
\dagger })}\int d\chi ^{\prime \dagger }d\chi ^{\prime }e^{\mathrm{tr}(\chi
^{\prime }\chi ^{\prime \dagger })}\langle O_{04}^{\dagger }O_{02}^{\dagger
}O_{03}O_{01}\rangle .
\end{equation}%
We then integrate out the auxiliary fields to get the correlation functions
of integrated operators%
\begin{equation}
\langle O_{4}^{\dagger }O_{2}^{\dagger }O_{3}O_{1}\rangle =\int d\chi
^{\dagger }d\chi e^{\mathrm{tr}(\chi \chi ^{\dagger })}G(\chi ,\chi
^{\dagger }).
\end{equation}%
If $O_{0}$ is identity operator, then $O$ is also identity operator.

\end{document}